\documentclass[prb,twocolumn,amsmath,amssymb,showpacs]{revtex4-1} %fjern 1 tal hvis gammel revtex4 version
\usepackage{graphics}
\usepackage{dcolumn}% Align table columns on decimal point
\usepackage{bm}% bold math
\usepackage{color}
\usepackage{epstopdf}
\usepackage{epsfig}
\newcommand{\down}{\downarrow}
\newcommand{\up}{\uparrow}
\newcommand{\spin}{\sigma}
\newcommand{\ospin}{\overline{\sigma}}
\newcommand{\kv}{{\bf k}}
\newcommand{\qv}{{\bf q}}
\newcommand{\Qv}{{\bf Q}}
\newcommand{\Nav}{\langle n \rangle}

\begin{document}

\title{Superconducting phase diagram of itinerant antiferromagnets}

\author{A. T. R\o mer,$^{1}$  I. Eremin,$^{2,3}$ P. J. Hirschfeld,$^{4}$ B. M. Andersen$^{1}$}
\affiliation{$^1$Niels Bohr Institute, University of Copenhagen, DK-2100 Copenhagen, Denmark\\
$^2$Institut f\"ur Theoretische Physik III, Ruhr-Universit\"at Bochum, D-44801 Bochum, Germany\\
$^3$National University of Science and Technology "MISiS", 119049 Moscow,
Russian Federation\\
$^4$ Department of Physics, University of Florida, Gainesville, USA
%$^4$ National University of Science and Technology ‘MISiS’, 119049 Moscow, Russia
}
\begin{abstract}

We study the phase diagram of the Hubbard model in the weak-coupling limit for coexisting spin-density-wave order and spin-fluctuation-mediated superconductivity. Both longitudinal and transverse spin fluctuations contribute significantly to the effective interaction potential, which creates Cooper pairs of the quasi-particles of the antiferromagnetic metallic state. We find a dominant $d_{x^2-y^2}$-wave solution in both electron- and hole-doped cases.  In the quasi-spin triplet channel, the longitudinal fluctuations give rise to an effective attraction
supporting a $p$-wave gap, but are overcome by repulsive contributions from the transverse fluctuations which disfavor $p$-wave pairing compared to $d_{x^2-y^2}$. The sub-leading pair instability is found to be in the $g$-wave channel, but complex admixtures of $d$ and $g$ are not energetically favored since their nodal structures coincide.
Inclusion of interband pairing, in which each fermion in the Cooper pair belongs to a different spin-density-wave band, is considered for a range of electron dopings in the regime of well-developed magnetic order. We demonstrate that these interband pairing gaps, which are non-zero in the magnetic state, must have the same parity under inversion as the normal intraband gaps. The self-consistent solution to the full system of five coupled gap equations give intraband and interband pairing gaps of $d_{x^2-y^2}$ structure and similar gap magnitude. In conclusion, the $d_{x^2-y^2}$ gap dominates for both hole and electron doping inside the spin-density-wave phase.
\end{abstract}


\maketitle

\section{INTRODUCTION}

Pairing of electrons by exchange of spin fluctuations is a popular paradigm proposed for unconventional superconductivity including Fe-based superconductors, cuprates, and heavy fermion systems. Since many of these systems exhibit an ordered magnetic phase coexisting with, or in close proximity to, the superconducting phase, a small number of studies have addressed the subsidiary problem of  pairing of quasi-particles in the symmetry broken spin-density-wave (SDW) phase.  While this problem has a long history, recent experimental and theoretical developments have led a number of authors to revisit it.\cite{IsmerPRL10,Schmiedt14,WenyaNJP}

For the simplest case of a doped one-band Hubbard model with standard commensurate $(\pi,\pi)$ ordering, the existence of pairing and its consequences for the symmetry of the superconducting order parameter was initially investigated  in a seminal paper by Schrieffer {\it et al.},~\cite{Schrieffer89} where the effective pairing interaction was obtained within the random phase approximation (RPA) arising from longitudinal spin fluctuations in the magnetically ordered phase.  These authors neglected  the contribution to pairing from the transverse spin fluctuations corresponding to the Goldstone mode of the spin symmetry broken state, arguing that  while such modes lead to a divergent contribution to the spin susceptibility at the ordering vector ${\bf Q}$,  the  coherence factors of the SDW phase screen the bare electron-electron interaction vertex, which therefore vanishes at ${\bf Q}$.
Soon after, Frenkel and Hanke~\cite{Frenkel90} showed that the transverse fluctuations do contribute to the pairing interaction in the same order as the longitudinal fluctuations; the divergence of the transverse spin susceptibility is eliminated by the coherence factors, but a residual constant interaction remains.

For electron-doped cuprates, the one-band Hubbard model seems to provide a reasonable minimal model since the doped electrons reside primarily on the copper sites.   Furthermore, calculations of band parameters for the electron-doped cuprates point to the fact that the Coulomb interaction is smaller than the bandwidth,~\cite{review-greene} in contrast to their  hole-doped counterparts.  In consequence, the mean field treatment of the SDW order works quite well for the normal state properties of the electron-doped systems. In particular, ARPES reports show a Fermi surface evolution upon increased electron doping which agrees well with the band reconstruction of the one-band Hubbard model due to commensurate $(\pi,\pi)$ order.  Upon electron doping, the Fermi surface consists of electron pockets around $(\pi,0)$, $(0,\pi)$. Close to critical electron doping, the emergence of hole pockets at (or close to) the Fermi level around $(\frac{\pi}{2},\frac{\pi}{2})$, $(-\frac{\pi}{2},\frac{\pi}{2})$ may occur.~\cite{review-greene}

Theoretically, the study of spin-fluctuation pairing for the electron-doped cuprates has been addressed mostly in the paramagnetic phase,~\cite{Manske00,Khodel04,Yoshimura,Kyung03,Parker08,Guinea,Astrid2015} where the mechanism was found to give rise to a gap with $d_{x^2-y^2}$ symmetry with strong non-monotonic features as a function of momentum as a result of Fermi surface intersections with the magnetic zone boundary, the so-called hot spots. At these positions the pairing becomes particularly pronounced. This behavior was found to agree qualitatively with ARPES~\cite{Matsui05} and Raman~\cite{Blumberg02} observations. In Ref.~\onlinecite{Guinea} the possibility for a cross-over to $d_{xy}$ symmetry was found at large dopings. In a later work, Krotkov and Chubukov~\cite{Krotkov} studied the spin-mediated pairing gap close to the quantum critical point of antiferromagnetic (AF) order and found an anisotropic $d_{x^2-y^2}$ gap symmetry, but the anisotropic behavior was not related to the hot spots, as opposed to previous work. A few studies also addressed the coexistence of superconductivity and long-range AF order but treated the pairing phenomenologically, and also found a $d_{x^2-y^2}$-wave solution.~\cite{IsmerPRL10,Das06,Ting06}

For hole-doped cuprates, stronger interactions imply that application of a weak-coupling approach to pairing is somewhat less justified.
In addition, calculations of the spin wave spectrum suggest that the commensurate $(\pi,\pi)$ order in this case is not the ground state solution,~\cite{ChubukovFrenkel,RowePRB12} complicating the theoretical modelling. Spin-fluctuation-mediated pairing has, however, been extensively applied to study also hole-doped cuprates within the one-band Hubbard model. First, for the paramagnetic phase Scalapino {\it et al.}~\cite{Scalapino86} generalised the approach of Berk and Schrieffer~\cite{berkschrie} and found a dominant $d_{x^2-y^2}$-wave pairing instability. Later, the study was extended by various methods and numerical approaches~\cite{Hlubina,Chubukov,Markiewicz08,Raghu10,Eberlein14,Astrid2015} including discussions of the possibility of other superconducting pairing symmetries arising from spin fluctuations.
From a strong coupling approach, as derived from the $t-J$ model, analysis of the effective pairing at low hole doping~\cite{Luschersushkov} suggests a $d$-wave superconducting ground state in any coexistence phase.  This is consistent with rigorous perturbative weak-coupling calculations in the presence of weak density-wave order~\cite{Choetal} as well as a study of hole-doped cuprates in the coexistence phase.~\cite{Galitski}
In addition, a recent study of spin-fluctuation mediated superconductivity in the SDW ordered metal attacked the problem by analytical RPA calculations in the large magnetization (small pocket) limit~\cite{WenyaNJP} and found that interactions were dominated by longitudinal fluctuations on the electron-doped side, supporting a nodeless $d_{x^2-y^2}$ gap, whereas on the hole-doped side both longitudinal and transverse fluctuations support a nodal $d_{x^2-y^2}$ gap.

Recently, Lu {\it et al.}~\cite{Lu2014} studied the case of underdoped cuprates in a $t-J$ like model and came to a rather different conclusion.  In this paper, the coexistence of commensurate AF and superconductivity was investigated in a phenomenological model where the pairing interaction arises based on nearest-neighbor magnetic exchange neglecting the double occupancy constraint. In the coexistence phase, this study found the leading superconducting instability to be triplet $p$-wave in the case of hole doping, providing a potential explanation for recent photoemission measurements indicating a ``nodal gap", a state with a full gap near the usual positions of the $d_{x^2-y^2}$ gap nodes whose existence is well-established at higher dopings.\cite{razzoli,peng} It is worth noting, however,  that in the original strong-coupling study of unconventional superconductivity driven by the spin waves, studied within the $t-J$ model, the $d_{x^2-y^2}$-wave symmetry of the superconducting gap was found to be the only stable solution.~\cite{Sushkov,Sushkov2} We also note that there exist other potential explanations of the existence of the nodeless gap in the literature.~\cite{bazak,das14,zhou2015}

This controversy suggests the need for a better understanding of the phase diagram of the single-band Hubbard model within a single, reliable approximation scheme which can encompass paramagnetic, superconducting and coexistence phases, and which is capable of identifying the strength of pairing by both longitudinal and transverse spin fluctuations and charge fluctuations, and their relative importance for pairing across a large doping range, for different electronic structures, and for both weak and strong magnetism. Therefore, we extend the work of Ref.~\onlinecite{WenyaNJP} by a more complete, fully numerical solution of the problem to both confirm the analytical calculations and extend them to the larger phase diagram.

We address the question of how the pairing interactions and resulting gap symmetry in the one-band Hubbard model, treated within the full spin-fluctuation approach with self-consistently determined SDW order, evolve as a function of doping throughout the  phase diagram, including the coexistence dome of SDW order and superconductivity. We show that the $d_{x^2-y^2}$ solution is in fact the leading superconducting instability for all electron doping levels.
In the hole-doped case, we limit ourselves to small hole doping only and force the $(\pi,\pi)$ order to be stable by a suppression of additional intrapocket nesting contributions to the transverse spin susceptibility.  This approach also yields a coexistence phase with $d_{x^2-y^2}$ order.
Thus, the weak-coupling approach where pairing is mediated by spin fluctuations gives qualitatively different results than the "strong-coupling" approach of Ref.~\onlinecite{Lu2014} in the case of hole doping. We find that the sub-leading instability  is a singlet $g$-wave solution, which shares common nodes with the  $d_{x^2-y^2}$ solution along the zone diagonals, and has additional nodes along the momentum axes. In the hole-doped case where the Fermi pockets are located far away from the zone axis, $g$ and $d_{x^2-y^2}$ become nearly degenerate. However, since the two solutions share common nodes, there is no gain in condensation energy by e.g. a time-reversal-symmetry broken solution of the form $d+ig$, and the solution of the full gap equation in fact favors $d_{x^2-y^2}$ over $g$.

We find further that extended $s$-wave is always suppressed in the coexistence phase, in agreement with the findings in Ref.~\onlinecite{WenyaNJP} for small doping levels, and on the electron-doped side we do not encounter a leading triplet gap at any moderate doping away from half-filling. The spin-fluctuation-pairing mechanism becomes strongly suppressed above the critical doping for which long-range magnetic order vanishes, since the nesting conditions are rapidly weakening as the Fermi surface segments of the paramagnetic phase move apart.

In a recent study of the iron-based superconductors by Hinojosa {\it et al.}~\cite{Hinojosa} it was pointed out that additional interband gaps may develop in the coexistence phase due to pairing of two fermions residing on different bands. Such pairs are naively expected to be negligible because they involve fermionic states far from the Fermi energy and do not alone manifest a Cooper instability.  In this regard, these
interband pairs can be considered as "anomalous". However, they are non-zero in the SDW state due to a combined effect of the SDW gap and coupling to the intraband gaps (dubbed in the following as normal gaps). In the iron-based system, this effect is predicted to create a different superconducting phase which explicitly breaks time-reversal symmetry by development of chiral gaps. We find from a self-consistent treatment of the coupled gap equations that the one-band Hubbard model also supports substantial anomalous interband pairings. The structural factor of the anomalous gaps is $d_{x^2-y^2}$ as for the intraband gaps, but the phase structure of the intraband and interband gaps can be different. However, unlike the case of iron-based system,~\cite{Hinojosa} the chiral solution $d_{x^2-y^2}^{\rm intra}+e^{i\phi}d_{x^2-y^2}^{\rm inter}$ does not appear to be favored in our numerical calculations.

\section{MODEL AND METHOD}
\subsection{SDW mean field Hamiltonian}
The model is the one-band Hubbard Hamiltonian
\begin{equation}
 H=-\sum_{i,j,\spin}t_{i,j}c_{i\spin}^\dagger c_{j\spin}+U\sum_{i}n_{i\up}n_{i\down}-\mu\sum_{i,\spin} n_{i\spin},
\label{eq:HubbardHRealSpace}
\end{equation}
where $c_{i\spin}^\dagger$ creates an electron on site $i$ with spin $\spin$. The interaction term $U$ denotes the energy cost associated with having two electrons on the same site. In reciprocal space the Hamiltonian reads
\begin{equation}
 H=\sum_{k\spin}\epsilon_\kv c_{\kv\spin}^\dagger c_{\kv\spin}+\frac{U}{2N}\sum_{\kv,\kv',q}\sum_{\spin}c_{\kv'\spin}^\dagger c_{-\kv'+\qv\ospin}^\dagger c_{-\kv+\qv\ospin} c_{\kv\spin},
\end{equation}
with
\begin{equation}
\epsilon_\kv=-2t[\cos(k_x)+\cos(k_y)]-4t'\cos(k_x)\cos(k_y)-\mu.
\end{equation}
The parameter $-t$ is the energy gain corresponding to hopping between neighboring sites, and $-t'$ denotes the energy gain by hopping to next-nearest neighbor sites. The doping level of the system is controlled by changing the chemical potential $\mu$.
The interaction between two electrons is first treated in the Hartree-Fock approximation giving rise to AF ordering of the spins. We therefore consider the mean field Hamiltonian
\begin{equation}
H_{\rm SDW}=\sideset{}{'}\sum_\kv\sum_\spin
(c_{\kv\spin}^\dagger \quad c_{\kv+\Qv\spin}^\dagger)
 \Big(\begin{array}{cc}
  \epsilon_\kv & \spin W\\
  \spin W & \epsilon_{\kv+\Qv}
 \end{array}
\Big)
 \Big(\begin{array}{c}
c_{\kv\spin}\\
c_{\kv+\Qv\spin}
 \end{array}
\Big),
\end{equation}
where $ W=-\frac{U}{N}\sum_{\kv}[\langle c_{\kv+\Qv\up}^\dagger c_{\kv\up}\rangle-\langle c_{\kv+\Qv\down}^\dagger c_{\kv\down}\rangle]$ is the AF order parameter, and $\sum^\prime$ refers to summation over the reduced Brillouin zone only.
Diagonalization of the mean field Hamiltonian leads to the energy spectrum
$ E_\kv^{\alpha,\beta}=\epsilon_\kv^+\pm \sqrt{(\epsilon_\kv^-)^2+W^2},  \hspace{.2cm} \epsilon_\kv^\pm=\frac{\epsilon_\kv\pm\epsilon_{\kv+\Qv}}{2}.$
The magnetic gap equation is solved self-consistently given the hopping integrals $t=1$, $t'$, the Coulomb repulsion $U$ and the doping. A gapless Goldstone mode at $\Qv=(\pi,\pi)$ in the transverse spin channel is obtained automatically, since the magnetic gap equation is equivalent to the condition $1-U\rm Re\chi_0^{+-}(\Qv)=0$.~\cite{Schrieffer89}
\subsection{Superconducting pairing interactions}
Higher order interactions in $U$ generate superconductivity on top of the AF order through longitudinal and transverse spin
fluctuations, following the original proposals of Refs.~\onlinecite{Scalapino86,Schrieffer89}. Since $U$ scatters the bare electrons, the diagrammatics are performed in terms of the bare electron Green's functions, but the Cooper pairing takes place between the quasi-particles of the AF state. The quasi-particle operators are related to the bare electron operators by the transformation:
\begin{eqnarray}
 c_{\kv\spin}&=&u_\kv\alpha_{k\spin}+v_\kv\beta_{\kv\spin}, \\
 c_{\kv+\Qv\spin}&=&{\rm sign}(\spin)[v_\kv\alpha_{\kv\spin}-u_\kv\beta_{\kv\spin}].
\end{eqnarray}
Transverse and longitudinal spin fluctuations give rise to fundamentally different interactions. Inspection of the interaction vertex formulated in real space~\cite{Romer12} shows that the charge and longitudinal interaction vertices give rise to no spin flips, whereas the transverse interaction does. In the latter channel, we have the gapless Goldstone mode of the AF phase which gives rise to a divergent interaction potential between the bare electrons. However, when we consider pairing between the quasi-particles of the AF state, this divergence is removed by the coherence factors as noted in earlier works.~\cite{WenyaNJP,Frenkel90} In the case of pairing between opposite spin electrons, spin flip processes are possible and the effective interaction is mediated both by longitudinal and transverse spin fluctuations.
If pairing occurs between same spin electrons, only longitudinal spin fluctuations contribute and the pairing potential does not include the bare Coulomb repulsion $U$ since this acts only between opposite spin electrons. The interaction Hamiltonian is formulated in terms of the SDW quasi-particles, and in line with earlier work~\cite{Schrieffer89,WenyaNJP} we show the interactions in the longitudinal and transverse channel individually, with the transverse part of the interaction stated as a spin flip vertex explicitly,

\begin{widetext}
\begin{eqnarray}
 H_{c/z}&=&\frac{1}{4N}\sideset{}{'}\sum_{\kv\kv'\spin}\Gamma_{\kv,\kv'}^z(\alpha_{\kv'\spin}^\dagger\alpha_{-\kv'\ospin}^\dagger\alpha_{-\kv\ospin}\alpha_{\kv\spin}+
\beta_{\kv'\spin}^\dagger\beta_{-\kv'\ospin}^\dagger\beta_{-\kv\ospin}\beta_{\kv\spin})
+\tilde\Gamma_{\kv,\kv'}^z(\alpha_{\kv'\spin}^\dagger\alpha_{-\kv'\ospin}^\dagger\beta_{-\kv\ospin}\beta_{\kv\spin}+
\beta_{\kv'\spin}^\dagger\beta_{-\kv'\ospin}^\dagger\alpha_{-\kv\ospin}\alpha_{\kv\spin})\nonumber \\
&&\label{eq:Hc},\\
 H_{\pm}&=&-\frac{1}{2N}\sideset{}{'}\sum_{\kv\kv'\spin}\Gamma^{+-}_{\kv,\kv'}(\alpha_{\kv'\spin}^\dagger\alpha_{-\kv'\ospin}^\dagger\alpha_{-\kv\spin}\alpha_{\kv\ospin}+
\beta_{\kv'\spin}^\dagger\beta_{-\kv'\ospin}^\dagger\beta_{-\kv\spin}\beta_{\kv\ospin})
+\tilde\Gamma^{+-}_{\kv,\kv'}(\alpha_{\kv'\spin}^\dagger\alpha_{-\kv'\ospin}^\dagger\beta_{-\kv\spin}\beta_{\kv\ospin}+
\beta_{\kv'\spin}^\dagger\beta_{-\kv'\ospin}^\dagger\alpha_{-\kv\spin}\alpha_{\kv\ospin}),\nonumber\\
&& \\
 H_{c/z}^{\rm ss}&=&\frac{1}{2N}\sideset{}{'}\sum_{\kv\kv'\spin}\Gamma^{\rm ss}_{\kv,\kv'}(\alpha_{\kv'\spin}^\dagger\alpha_{-\kv'\spin}^\dagger\alpha_{-\kv\spin}\alpha_{\kv\spin}+
\beta_{\kv'\spin}^\dagger\beta_{-\kv'\spin}^\dagger\beta_{-\kv\spin}\beta_{\kv\spin})
+\tilde\Gamma^{\rm ss}_{\kv,\kv'}(\alpha_{\kv'\spin}^\dagger\alpha_{-\kv'\spin}^\dagger\beta_{-\kv\spin}\beta_{\kv\spin}+
\beta_{\kv'\spin}^\dagger\beta_{-\kv'\spin}^\dagger\alpha_{-\kv\spin}\alpha_{\kv\spin}), \nonumber \\
&& \label{eq:Hss}
\end{eqnarray}

with

\begin{eqnarray}
\Gamma^z_{\kv,\kv'}&=&[2U-V_c(\kv-\kv')]l^2(\kv,\kv')-[2U-V_c(\kv-\kv'+\Qv)]m^2(\kv,\kv') +V_z(\kv-\kv')l^2(\kv,\kv')-V_z(\kv-\kv'+\Qv)m^2(\kv,\kv') \nonumber, \\
&& \label{eq:GammaZ}\\
\Gamma^{+-}_{\kv,\kv'}&=&V_{+-}(\kv-\kv')n^2(\kv,\kv')-V_{+-}(\kv-\kv'+\Qv)p^2(\kv,\kv')\label{eq:GammaPM}, \\
&& \nonumber \\
\Gamma^{\rm ss}_{\kv,\kv'}&=&-V_{c}(\kv-\kv')l^2(\kv,\kv')-V_{c}(\kv-\kv'+\Qv)m^2(\kv,\kv')-V_{z}(\kv-\kv')l^2(\kv,\kv')-V_{z}(\kv-\kv'+\Qv)m^2(\kv,\kv').\label{eq:GammaSS}
\end{eqnarray}
\end{widetext}

The main ingredients in the pairing interactions are the spin and charge susceptibilities within the RPA approximation
\begin{equation}
V_c(\qv)=\frac{U^2\chi_0^{z}(\qv)}{1+U\chi_0^{z}(\qv)},
\end{equation}
\begin{equation}
V_z(\qv)=\frac{U^2\chi_0^{z}(\qv)}{1-U\chi_0^{z}(\qv)},
\end{equation}
\begin{equation}
V_{+-}(\qv)=\frac{U^2\chi_0^{+-}(\qv)}{1-U\chi_0^{+-}(\qv)},
\end{equation}
where the spin susceptibilities are defined by $\chi_0^{z}(\qv,\omega)=\frac{i}{2N}\int dt e^{i\omega t}\langle T S_\qv^z (t)S_{-\qv}^z\rangle $ and $\chi_0^{+-}= \frac{i}{2N}\int dt e^{i\omega t}\langle T S_\qv^+ (t)S_{-\qv}^-\rangle $.
As seen from Eqs.~(\ref{eq:GammaZ})-(\ref{eq:GammaSS}) the bare interaction vertices are modified by coherence factors of the SDW phase given by $ u_\mu^2(\kv,\kv')=\frac{1}{2}\Big(1+(-1)^\mu\frac{\epsilon_\kv^-\epsilon_{\kv'}^-+\nu_\mu W^2}{\sqrt{(\epsilon_\kv^-)^2+W^2}\sqrt{(\epsilon_{\kv'}^-)^2+W^2}}\Big)
$
 with $u_\mu^2=m^2,l^2,p^2,n^2$ and $\nu_\mu=(-1,1,1,-1)$.

% --------- modified ------------
The interaction Hamiltonians stated in Eqs.~(\ref{eq:Hc})-(\ref{eq:Hss}) are restricted to Cooper pairing between quasi-particles residing in the same band, i.e. $\langle\alpha_{\kv\spin}^\dagger \alpha_{-\kv\ospin}^\dagger\rangle$ and $\langle\beta_{\kv\spin}^\dagger \beta_{-\kv\ospin}^\dagger\rangle$. In section ~\ref{sec:anomalouspairing} we introduce an extended model which includes Cooper pairing between fermions residing on different pockets, i.e. mean fields of the form $\langle\alpha_{\kv\spin}^\dagger \beta_{-\kv\ospin}^\dagger\rangle$.
For now we restrict ourselves to normal intraband Cooper pairs and include pair scattering interactions within each band as well as between the bands. % ---
 The expression for the interband couplings between pairs of $(\kv,\kv')$ with $\kv$ residing on the $\alpha$ band and $\kv'$ on the $\beta$ band, which are dubbed $\tilde\Gamma_{\kv,\kv'}^z$, $\tilde\Gamma_{\kv,\kv'}^{+-}$ and $\tilde\Gamma_{\kv,\kv'}$ are obtained by interchanging the coherence factors $p^2(\kv,\kv') \leftrightarrow l^2(\kv,\kv')$ and $m^2(\kv,\kv') \leftrightarrow n^2(\kv,\kv')$ in the Eqs. (\ref{eq:GammaZ})-(\ref{eq:GammaSS}) in agreement with Refs.~\onlinecite{Schrieffer89,WenyaNJP}.

The presence of magnetic order breaks spin-rotational
symmetry, but inversion symmetry is preserved. This
allows us to express the superconducting gap in an even and odd parity form which corresponds to a quasi-spin
singlet and  a quasi-spin triplet gap, respectively. We label the gaps by the superscript "s" for even parity (singlet) and superscript "t" for the odd parity (triplet) gap, i.e. $\Delta_{\kv}^{\alpha (\rm s/t)}=\langle \alpha_{-\kv\down}\alpha_{\kv\up}\rangle\mp\langle \alpha_{-\kv\up}\alpha_{\kv\down}\rangle$. As we show later when discussing the interband (anomalous) pairing terms, it
is more useful to classify the superconducting gaps in the SDW background by
parity rather than by spin quantum numbers. In this regard calling Cooper pairing spin singlet or spin triplet in the SDW background actually refers to the even or odd parity of the wave function, respectively.
 
In the derivation of the superconducting gap equations the splitting of these two channels leads to the symmetrization/antisymmetrization of the pairing potential in the quasi-spin singlet/triplet gap equation
\begin{eqnarray}
\tiny
 \Delta^{\alpha, \rm (s/t)}_\kv
&=&-\frac{1}{8N}\sideset{}{'}\sum_{\kv'}\Big[\Gamma^{\rm (s/t)}_{\kv,\kv'}\frac{\Delta^{\alpha,\rm (s/t)}_{\kv'}}{\Omega_{\kv'}^{\alpha,\rm (s/t)}}\tanh\Big(\frac{\Omega_{\kv'}^{\alpha,\rm (s/t)}}{2k_BT}\Big)\nonumber \\
&&\hspace{1.52cm}+\tilde{\Gamma}^{\rm (s/t)}_{\kv,\kv'}\frac{\Delta^{\beta\rm (s/t)}_{\kv'}}{\Omega_{\kv'}^{\beta\rm (s/t)}}\tanh\Big(\frac{\Omega_{\kv'}^{\beta\rm (s/t)}}{2k_BT}\Big)
\Big]\nonumber,\\
\label{eq:SCGapEquation}
\end{eqnarray}
and similarly for $\Delta^{\beta,\rm (s/t)}_\kv$ by interchanging $\alpha \leftrightarrow \beta$. In the case of opposite spin interaction
the effective pairing interactions for the singlet and triplet channel are given by
\begin{eqnarray}
\Gamma^{\rm (s)}_{\kv,\kv'}&=&(\Gamma^z_{\kv,\kv'}+2\Gamma^{+-}_{\kv,\kv'})+(\Gamma^z_{-\kv,\kv'}+2\Gamma^{+-}_{-\kv,\kv'}), \label{eq:symPots}\\
\Gamma^{\rm (t)}_{\kv,\kv'}&=&(\Gamma^z_{\kv,\kv'}-2\Gamma^{+-}_{\kv,\kv'})-(\Gamma^z_{-\kv,\kv'}-2\Gamma^{+-}_{-\kv,\kv'}). \label{eq:symPott}
\end{eqnarray}
for the intraband contributions and equivalent expressions for the interband, $\tilde \Gamma$, contributions. Similarly, for same spin electron interactions, the singlet and triplet potential is obtained directly by a symmetrization/antisymmetrization of the potential stated in Eq.~(\ref{eq:GammaSS}).

The strength of the pairings is calculated by evaluating the real part of the RPA susceptibilities at zero energy. Note that due to broken spin rotation symmetry there is a difference between $\chi_0^z$ and $\chi_0^{+-}$. The pairing vertex $V_{+-}(\qv)$ diverges at $\qv=\Qv$ due to the Goldstone mode in the transverse channel. This divergence is removed, however, by the coherence factor $p^2(\kv,\kv')$ of the SDW phase, as discussed above.

In the paramagnetic phase where $W=0$, pairing takes place between the bare electrons and the gap equation then reduces to
\begin{eqnarray}
 \Delta^{\rm (s/t)}_\kv&=&
-\frac{1}{4N}\sum_{\kv'}[V^{\rm (s/t)}_{\kv,\kv'}\pm V^{\rm (s/t)}_{-\kv,\kv'}]\frac{\Delta^{\rm (s/t)}_{\kv'}}{E^{\rm (s/t)}_{\kv'}}\tanh\Big(\frac{E^{\rm (s/t)}_{\kv'}}{2k_BT}\Big),\nonumber\\
\label{eq:SCGapEquationNS}
\end{eqnarray}
with $E^{\rm (s)/(t)}_\kv=\sqrt{\xi_{\kv}^2+|\Delta^{\rm (s/t)}_{\kv}|^2}$. The effective pairings are given by
\begin{eqnarray}
V^{\rm (s/t)}_{\kv,\kv'}&=&U+\frac{1}{2}[V_z(\kv-\kv')-V_c(\kv-\kv')]\pm V_{+-}(\kv-\kv').\nonumber\\
 \label{eq:PMpot}
\end{eqnarray}

\subsection{Gap symmetries in the SDW phase}
In the SDW phase, the effective pairings contain Umklapp terms, which are the terms in Eqs.~(\ref{eq:GammaZ})-(\ref{eq:GammaSS}) containing the argument $\kv-\kv'+\Qv$. Due to these terms, there are attractive pair scatterings $(\kv,\kv')$ on the reconstructed Fermi surface. This contrasts with the situation in the paramagnetic phase, where the spin-fluctuation part of singlet potential is always purely repulsive for all pairs of $(\kv,\kv')$, as seen from Eq.~(\ref{eq:PMpot}).
However, despite the presence of a partially attractive potential, a conventional $s$-wave superconducting gap is not possible. This is due to a symmetry constraint on the pairing potentials, which must obey $\Gamma_{\kv,\kv'}=-\Gamma_{\kv+\Qv,\kv'}=-\Gamma_{\kv,\kv'+\Qv}$, a property that is fulfilled for both the longitudinal and transverse pairing interactions.
This symmetry also carries over to the superconducting gap, which must satisfy
\begin{equation}
\Delta_\kv=-\Delta_{\kv+\Qv}.
\label{eq:Dsym}
\end{equation}
For the two dimensional square lattice, the gap solutions can be classified according to the five irreducible representations of the $D_{4h}$ group that are even under reflection through the horizontal plane, i.e. extended $s$-wave ($s^*$), $d_{x^2-y^2}$, $d_{xy}$, $g$ and the triplet solution, $p_x/p_y$, which is doubly degenerate.
First we consider whether these solutions all comply with the additional symmetry constraint, Eq.~(\ref{eq:Dsym}) present in the SDW phase. This disqualifies the $d_{xy}$ solution and as a result we consider the leading gap symmetries from among the set
\begin{eqnarray}
A_{1g}:& \quad& s^*=\cos(k_x)+\cos(k_y),\label{eq:exts} \\
B_{1g}:& ~&   d_{x^2-y^2}=\cos(k_x)-\cos(k_y) \label{eq:d2}, \\
A_{2g}:& ~&    g=[\cos(k_x)-\cos(k_y) ]\sin(k_x)\sin(k_y), \label{eq:gxy}\\
E_{u}:& ~&    p_{x}=\sin(k_x), \quad  p_{y}=\sin(k_y) \label{eq:p}.
\end{eqnarray}
We emphasize that these basis functions are only the lowest order functions corresponding to the given irreducible representations.
In the construction of higher order solutions in the SDW phase, it is important to note that an odd number of the above basis functions must be multiplied in order for the resulting higher order gap function to comply with Eq.~(\ref{eq:Dsym}).
Further allowed higher order solutions can also be achieved
by a multiplication of the $d_{xy}$ basis function [$\sin (k_x) \sin (k_y)$] with any
$A_{1g}$, $B_{1g}$ or $E_{u}$ basis  functions.
A relevant triplet solution
is constructed by a multiplication of the triplet $p$-wave with the lowest order A$_{1g}$ and B$_{1g}$ basis functions,
\begin{eqnarray}
p^\prime_{x}&=&[\cos(k_x)-\cos(k_y)][\cos(k_x)+\cos(k_y)]\sin(k_x), \label{eq:extf} \nonumber \\
\end{eqnarray}
which for tetragonal symmetry belongs to the same $E_u$ symmetry representation as the original $p$-wave solution. Therefore, we dub this state $p^\prime$-wave. Note that in the literature on Sr$_2$RuO$_4$ this function is sometimes called $f_{x^2-y^2}$-wave.~\cite{EreminEPL02}
We find that this is the leading solution among the triplet solutions for any doping. However, as we shall see below, it does not dominate over the singlet solutions.
\begin{figure}[t!]
\centering
 	\includegraphics[angle=0,width=0.95\linewidth]{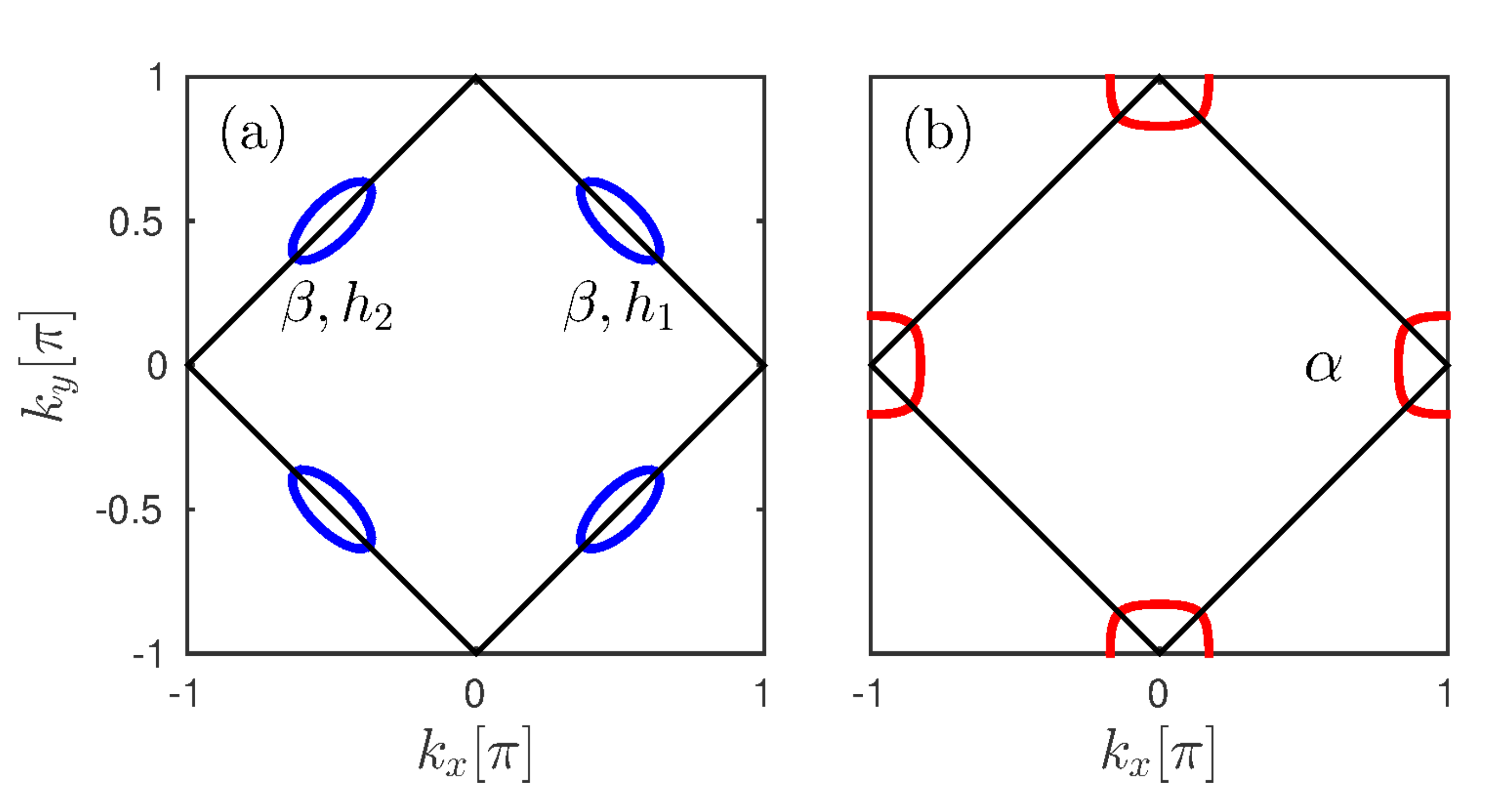}
 	\caption{(color online) Fermi surfaces in the case of (a) small hole doping, $\Nav =0.96$  and (b) small electron doping $\Nav=1.05$ for $U=3$ and $t'=-0.35$. }
\label{fig:Fermisurf}
\end{figure}

The consequences of a $d_{x^2-y^2}$ solution are manifested very differently depending on the Fermi surface geometry. In the presence of only electron pockets, as in Fig.~\ref{fig:Fermisurf}(b), it is nodeless at the Fermi surface, as opposed to the other solutions, which all display nodes.
If both electron and hole pockets are present at the Fermi surface as in Fig.~\ref{fig:solutionsLGE}(e), the $d_{x^2-y^2}$ solution has nodes only at the hole pockets, whereas
the extended $s$-wave as well as the $g$-wave solutions exhibit nodes at both types of pockets.
Na\"ively, we might expect the preferred solution to have the minimum number of nodes at the Fermi surface. In this respect, the $d_{x^2-y^2}$ solution clearly wins, but ultimately the leading solution relies on a detailed investigation of the structure of the pairing interaction in the SDW phase and the solution to the full gap equation. Finally, when only hole pockets are present at the Fermi surface as shown in Fig.~\ref{fig:Fermisurf}(a), the $p_x$ solution becomes nodeless, whereas none of the singlet gap symmetries will provide nodeless superconductivity. As we show below, the  $p_x$ solution is supported from the structure of the longitudinal fluctuations, but  becomes overall less favorable compared to $d_{x^2-y^2}$  due to an effective repulsion arising from the transverse spin fluctuations.

\subsection{Linearized gap equation in the SDW phase}
In order to determine the leading and sub-leading instabilities, we linearize the full gap equation stated in Eq.~(\ref{eq:SCGapEquation}) to obtain the eigenvalue problem
\begin{equation}
-\frac{1}{4(2\pi)^2} M_{\kv,\kv'}\Delta_{\kv'}=\lambda\Delta_{\kv'},
\label{eq:lge}
\end{equation}
The eigenvector of this equation
\begin{eqnarray}
 \Delta_{\kv}=\left[ \begin{array}{c}
                                                             \Delta_\kv^{\alpha} \\
                                                            \quad \Delta_\kv^{\beta_{h_1}}\\
                                                            \quad \Delta_\kv^{\beta_{h_2}}\\
                                                            \end{array}\right].
\end{eqnarray}
is a function of momentum $\kv$. The momentum is located either on the electron pocket around $(\pi,0)$, which we denote by $\alpha$, or on the two hole pockets, denoted by~$\beta_{h_1}/\beta_{h_2}$, around $(\frac{\pi}{2},\frac{\pi}{2})$/$(-\frac{\pi}{2},\frac{\pi}{2})$, respectively, see Fig.~\ref{fig:Fermisurf}.
We solve the eigenvalue problem Eq.~(\ref{eq:lge}) for $\kv$ and $\kv'$ on these three closed pockets, which is equivalent to solving it in the whole magnetic Brillouin zone.
All intraband and interband interactions are included in $M_{\kv,\kv'}$:
\begin{equation}
\left[ \begin{array}{c c c}
                           \Gamma^{\alpha\alpha}_{\bf \kv,\kv'}l^\alpha_{\kv'}/|v^\alpha_{\kv'}| & \tilde{\Gamma}^{\alpha \beta_{h_1}}_{\bf \kv,\kv'}l^{\beta}_{\kv'}/|v^\beta_{\kv'}| &\tilde{\Gamma}^{\alpha \beta_{h_2}}_{\bf \kv,\kv'} l^{\beta}_{\kv'}/|v^\beta_{\kv'}|\\
                           \tilde{\Gamma}^{\beta_{h_1}\alpha}_{\bf \kv,\kv'}l^\alpha_{\kv'}/|v^\alpha_{\kv'}| & \Gamma^{\beta_{h_1}\beta_{h_1}}_{\bf \kv,\kv'} l^{\beta}_{\kv'}/|v^\beta_{\kv'}|&\Gamma^{\beta_{h_1}\beta_{h_2}}_{\bf \kv,\kv'} l^{\beta}_{\kv'}/|v^\beta_{\kv'}|\\
                           \tilde{\Gamma}^{\beta_{h_2}\alpha}_{\bf \kv,\kv'}l^\alpha_{\kv'}/|v^\alpha_{\kv'}| & \Gamma^{\beta_{h_2}\beta_{h_1}}_{\bf \kv,\kv'}l^{\beta}_{\kv'}/|v^\beta_{\kv'}| &\Gamma^{\beta_{h_2}\beta_{h_2}}_{\bf \kv,\kv'} l^{\beta}_{\kv'}/|v^\beta_{\kv'}|
                                                            \end{array}\right]. \nonumber
\end{equation}
Here, $l_{\kv'}$ denotes the length of the Fermi surface line segment represented by the point $\kv'$ and $v_{\kv'}$ is the Fermi velocity at $\kv'$.
The intraband ($\Gamma_{\kv \kv'}$) and interband ($\tilde{\Gamma}_{\kv \kv'}$) pairings must be expressed in the singlet and triplet symmetrized versions
in accordance with Eqs.~(\ref{eq:symPots})-(\ref{eq:symPott}).
The largest eigenvalue, $\lambda$, gives the leading instability since it corresponds to the largest superconducting critical temperature and the symmetry of the gap is given by the corresponding eigenvector.

The linearized gap equation does not allow for a determination of complex gap solutions, which are time-reversal-symmetry-broken (TRSB) solutions, since higher order interactions in $\Delta_\kv$ are removed in the linearization process. Therefore, we have additionally solved the full nonlinear gap equation as given  in Eq.~(\ref{eq:SCGapEquation}). This allows us to compare the real solutions and TRSB solutions energetically. The latter type of solutions arise naturally in situations where the pairing potential allows for two degenerate eigenfunctions of $M_{\kv,\kv'}$. Therefore, they are likely to appear in the triplet channel, where all solutions to the linearized gap equation are two-fold degenerate. Since the linearized equation in fact also exhibits accidentally degenerate solutions in the singlet channel upon hole doping, this opens the possibility of TRSB states in this case. The prospect of observation of transitions to TRSB states as a function of doping or other control parameters, potentially the first observation of this kind, has recently been the subject of considerable attention in Fe-based superconductors.~\cite{lee09,platt12,maiti13}

\subsection{Anomalous pairing gaps}
\label{sec:anomalouspairing}
So far, we have restricted the discussion to Cooper pairing between quasi-particles residing on the same band, which are Cooper pairs of the form $\langle\alpha_{\kv\spin}^\dagger \alpha_{-\kv\ospin}^\dagger\rangle$ and $\langle\beta_{\kv\spin}^\dagger \beta_{-\kv\ospin}^\dagger\rangle$. Now we introduce an extension of the model to also include anomalous pairings between fermions belonging to different pockets, i.e. $\langle\alpha_{\kv\spin}^\dagger \beta_{-\kv\ospin}^\dagger\rangle$. The complete interaction Hamiltonian includes pair scattering processes between normal and anomalous gaps, as well as scatterings between anomalous gaps. 

Anomalous pairs of the form $\langle\alpha_{\kv\spin}^\dagger \beta_{-\kv\ospin}^\dagger\rangle$ involve fermions far from the Fermi level, since the two bands are gapped by $|E^\alpha_\kv-E^\beta_\kv| \geq 2W$. Nevertheless, the anomalous gaps become sizeable due to the coupling to the normal intraband pairs and exist only in the SDW phase since the coupling between normal and anomalous gaps are proportional to $W$. A detailed analysis of the interaction Hamiltonian reveals that {\it even} parity intraband gaps, i.e. $\langle \alpha_{\kv\up}^\dagger \alpha_{-\kv\down}^\dagger \rangle -\langle \alpha_{\kv\down}^\dagger \alpha_{-\kv\up}^\dagger \rangle$ couple to {\it even} parity anomalous interband gaps. In the iron-pnictide study of Ref.~\onlinecite{Hinojosa} the reported anomalous gap was dubbed quasi-spin triplet and coupled to an even parity singlet intraband gap. However, in that case the triplet gap is actually of even parity, which is allowed due to the band index. As a matter of fact, in the SDW background it is more useful to classify the superconducting gaps by parity rather than by spin quantum numbers, as spin-rotational symmetry is explicitly broken. Thus, in our case as well as in Ref.~\onlinecite{Hinojosa} one has an even-parity normal intraband gap coupling to even-parity anomalous pairing contributions. As we shall see below, calculations of the normal intraband gaps reveal an even-parity $d_{x^2-y^2}$-wave solution.
Because of this, and also as a result of the structure of the interaction Hamiltonian, we restrict ourselves to the even-parity channel which we label by a quasi-spin singlet index, $s$. In the even-parity channel, the mean-field interaction Hamiltonian takes the form

\begin{widetext}
\begin{eqnarray}
 H_{\Delta}&=&-\sideset{}{'}\sum_{\kv}\Big[\Delta_{\alpha\alpha}^s(\kv)\alpha_{-\kv\down}\alpha_{\kv\up} +\Delta_{\beta\beta}^s(\kv)\beta_{-\kv\down}\beta_{\kv\up} +\Delta_{\alpha\beta}^s(\kv)[\alpha_{-\kv\up}\beta_{\kv\down}-\alpha_{-\kv\down}\beta_{\kv\up}+\beta_{-\kv\up}\alpha_{\kv\down}-\beta_{-\kv\down}\alpha_{\kv\up}] \nonumber\\
 && \nonumber \\
 && \hspace{.8cm}+\Delta_{\alpha\beta\up\down}^s(\kv)\alpha_{-\kv\down}\beta_{\kv\up} -\Delta_{\alpha\beta\down\up}^s(\kv)\beta_{-\kv\down}\alpha_{\kv\up} + h.c.
\Big],\nonumber\\
\label{eq:HMFfull}
\end{eqnarray}
with the mean fields
\begin{eqnarray}
  \Delta_{\alpha\alpha}^s(\kv)&=&-\frac{1}{8N}\sideset{}{'}\sum_{\kv'}
  i\sigma_{\gamma\delta}^y[\Gamma^{\rm s}_{\kv,\kv'}  \langle\alpha_{\kv' \gamma }^\dagger \alpha_{-\kv' \delta }^\dagger  \rangle +
 \tilde\Gamma_{\kv,\kv' }^{\rm s}\langle\beta_{\kv' \gamma }^\dagger \beta_{-\kv' \delta }^\dagger  \rangle]
 +i\sigma_{\gamma\delta}^y\Gamma^{\alpha\beta,s}_{\kv,\kv'} \langle \alpha_{\kv' \delta }^\dagger \beta_{-\kv' \gamma}^\dagger +
 \beta_{\kv' \delta }^\dagger \alpha_{-\kv' \gamma}^\dagger \rangle,  \label{eq:Daa}\\
 &&\nonumber\\
 \Delta_{\beta\beta}^s(\kv)&=&-\frac{1}{8N}\sideset{}{'}\sum_{\kv'}
  i\sigma_{\gamma\delta}^y[\Gamma^{\rm s}_{\kv,\kv'}\langle \beta_{\kv' \gamma}^\dagger \beta_{-\kv' \delta}^\dagger \rangle +
\tilde\Gamma^{\rm s}_{\kv,\kv'}\langle \alpha_{\kv' \gamma }^\dagger \alpha_{-\kv' \delta }^\dagger  \rangle] -
   i\sigma_{\gamma\delta}^y \Gamma^{\alpha \beta,s}_{\kv,\kv'}
   \langle \alpha_{\kv' \delta }^\dagger \beta_{-\kv' \gamma}^\dagger +
 \beta_{\kv' \delta }^\dagger \alpha_{-\kv' \gamma}^\dagger \rangle,
 \label{eq:Dbb}\\
 &&  \nonumber  \\
 \Delta_{\alpha\beta}^s(\kv)&=&-\frac{1}{8N}\sideset{}{'}\sum_{\kv'}
 i\sigma_{\gamma\delta}^y \Gamma^{\alpha \beta,s}_{\kv,\kv'}\langle \alpha_{\kv' \gamma }^\dagger \alpha_{-\kv' \delta }^\dagger
- \beta_{\kv' \gamma }^\dagger \beta_{-\kv' \delta }^\dagger \rangle, \label{eq:gapano}  \\
 \Delta_{\alpha\beta \up\down}^s(\kv)&=&-\frac{1}{4N}\sideset{}{'}\sum_{\kv'} \Gamma^{\alpha\beta\alpha\beta,1,s}_{\kv,\kv'} \langle \alpha_{\kv' \up}^\dagger  \beta_{-\kv' \down}^\dagger  \rangle
 +\Gamma^{\alpha\beta\alpha\beta,2,s}_{\kv,\kv'} \langle \beta_{\kv' \up}^\dagger
 \alpha_{-\kv' \down}^\dagger  \rangle , \\
 \Delta_{\alpha\beta \down\up}^s(\kv)&=&-\frac{1}{4N}\sideset{}{'}\sum_{\kv'} \Gamma^{\alpha\beta\alpha\beta,1,s}_{\kv,\kv'} \langle \alpha_{\kv' \down}^\dagger
 \beta_{-\kv' \up}^\dagger  \rangle +\Gamma^{\alpha\beta\alpha\beta,2,s}_{\kv,\kv'} \langle \beta_{\kv' \down}^\dagger
 \alpha_{-\kv' \up}^\dagger  \rangle.
 \label{eq:gapanodownup}
\end{eqnarray}
In these gap equations we have introduced three new spin-fluctuation-mediated pairing interactions
\begin{eqnarray}
\Gamma^{\alpha \beta}_{\kv,\kv'}&=& \mp [V_{lo}(\kv-\kv')\nu(\kv,\kv')+V_{lo}(\kv-\kv'+\Qv)\mu(\kv,\kv')] -[ 2V_{+-}(\kv-\kv')\mu(\kv,\kv') + 2V_{+-}(\kv-\kv'+\Qv)\nu(\kv,\kv')], \nonumber \\
&& \label{eq:pairingAAAB} \\
\Gamma^{\alpha\beta\alpha\beta,1}_{\kv,\kv'}&=& -[V_{lo}(\kv-\kv')p^2(\kv,\kv')+V_{lo}(\kv-\kv'+\Qv)n^2(\kv,\kv')+2V_{+-}(\kv+\kv')n^2(\kv,\kv')+2V_{+-}(\kv+\kv'+\Qv)p^2(\kv,\kv')],\nonumber \\
&&\\
 \Gamma^{\alpha\beta\alpha\beta,2}_{\kv,\kv'}&=& V_{lo}(\kv-\kv')l^2(\kv,\kv')+V_{lo}(\kv-\kv'+\Qv)m^2(\kv,\kv') + 2V_{+-}(\kv+\kv')m^2(\kv,\kv')+2V_{+-}(\kv+\kv'+\Qv)l^2(\kv,\kv')],\nonumber\\
\label{eq:GammaABAB2}
 \end{eqnarray}
where
$ V_{lo}(\qv)=2U-V_c(\qv)+V_z(\qv)$, and $\mp$ in Eq.~(\ref{eq:pairingAAAB}) refers to the quasi-spin singlet and triplet, respectively. In Eqs.~(\ref{eq:Daa})-(\ref{eq:gapanodownup}) the superscript $s$ refers to the symmetrized potentials, e.g. $\Gamma^{\alpha\beta,s}_{\kv,\kv'}=\Gamma^{\alpha\beta}_{\kv,\kv'}+\Gamma^{\alpha\beta}_{-\kv,\kv'}$. This ensures that all mean fields are of even parity.

We have introduced two new coherence factors, which are proportional to the magnetic order parameter $W$
\begin{eqnarray}
  \mu(\kv,\kv')&=&  \frac{W}{2}\Big(\frac{\epsilon_{\kv'}^-+\epsilon_\kv^-}{E_\kv^-E_{\kv'}^-}\Big),\\
\nu(\kv,\kv')&=& \frac{W}{2}\Big(\frac{\epsilon_{\kv'}^--\epsilon_\kv^-}{E_\kv^-E_{\kv'}^-}\Big).
\end{eqnarray}

\end{widetext}
Note that the coupling between the normal intraband gaps and the anomalous pairings occurs due to the pairing stated in Eq.~(\ref{eq:pairingAAAB}) which is proportional to the SDW order parameter.

\begin{figure}[t!]
\centering
 	\includegraphics[angle=0,width=0.95\linewidth]{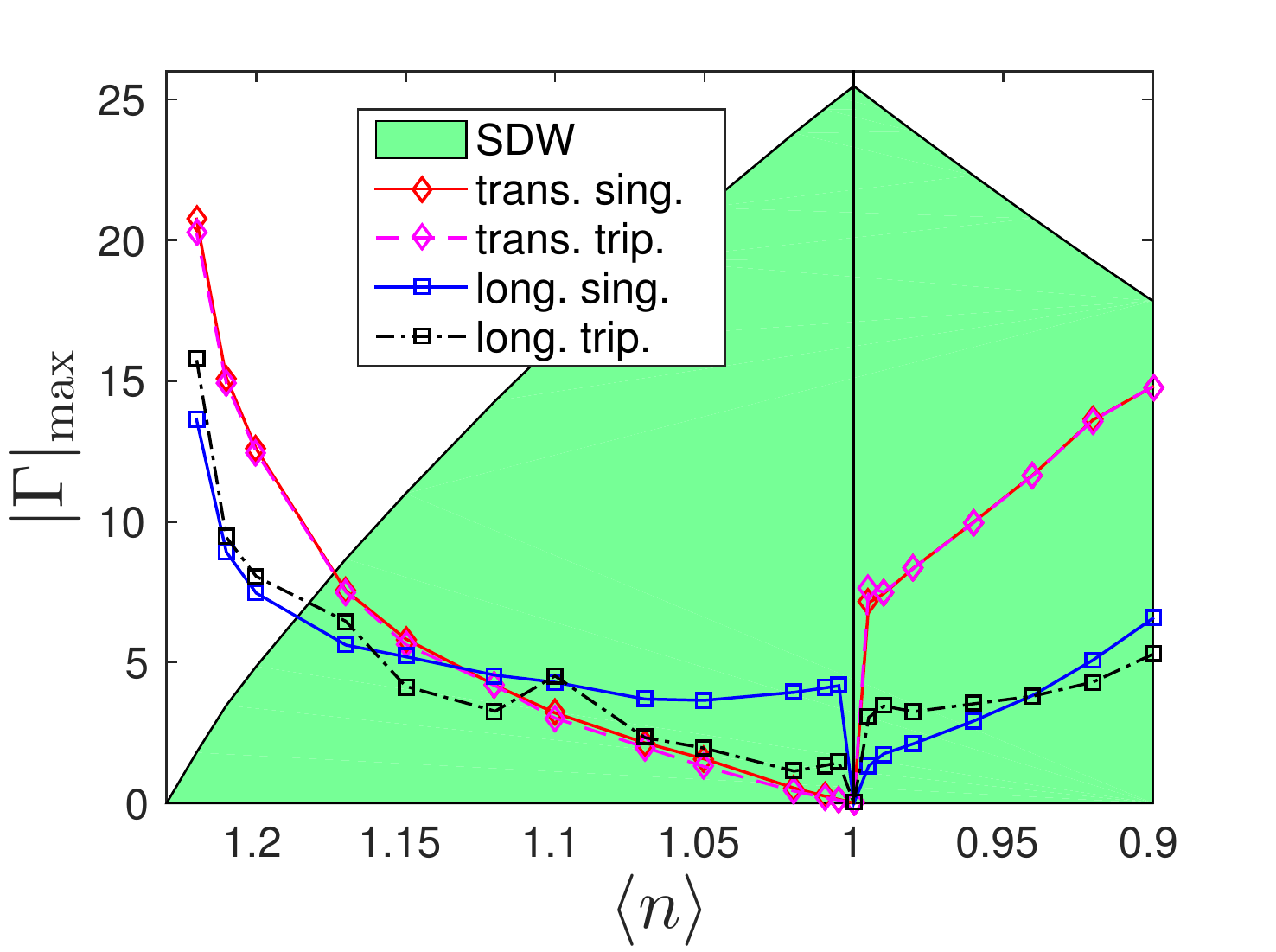}
 	\caption{(color online) Maximum value of the pairing contributions from longitudinal and transverse spin fluctuations in the singlet and triplet channel, $\Gamma_{\kv,\kv'}^{s/t}$, in the SDW phase as a function of doping, for the parameters $t'=-0.35$ and $U=3$.}
\label{fig:potphasediagram}
\end{figure}

\section{RESULTS}
\label{section_res}
In the next sections, we discuss the results of spin-fluctuation-mediated pairing in the SDW phase.
First, we restrict the model to normal intraband Cooper pairs only. In section~\ref{sec:pairingstrength} we study how the effective interactions mediated by transverse and longitudinal spin fluctuations evolve as a function of doping.
Thereafter, we turn to the implications for the evolution of different gap symmetry solutions in section~\ref{sec:lgeresults}. We focus on electron doping, since this is the most relevant regime for a coexistence phase of commensurate AF order and superconductivity. In the case of hole doping, we restrict the investigation to small hole doping levels below 10 \%.
Lastly, we provide a discussion of the extended model, where anomalous pairings are included, in section~\ref{sec:anomalouspairingresults}.
\subsection{Pairing interactions from longitudinal and transverse fluctuations}
\label{sec:pairingstrength}
First we focus on the strength and structure of the pairing potentials between opposite spin electrons arising from longitudinal and transverse spin fluctuations for hole and electron doping.
In Fig.~\ref{fig:potphasediagram} we plot the value of the pair scattering $\Gamma_{\kv,\kv'}^{s/t}$ for which the pairing strength is maximal for any pair of momenta $(\kv,\kv ')$ on the Fermi surface. In the low doping limit, we observe that the interaction through transverse fluctuations is much stronger for hole doping than for electron doping. For interactions in the longitudinal channel, the situation is opposite, since in this case the pairings are strongest for small electron doping compared to small hole doping.

Generally, for the hole-doped system, transverse fluctuations are quantitatively stronger than the longitudinal fluctuations.
For electron dopings, the longitudinal fluctuations dominate close to half filling, whereas both types of pairings contribute more equally at larger doping levels.
In fact, the strength of the pairing interactions in both channels builds up towards the critical electron doping where the AF order disappears. This na\"ively suggests an increase in superconductivity close to critical electron doping.
However, a closer inspection of the $(\kv,\kv')$ dependence of the pairing reveals that it
develops a sign change for a part of the intrapocket potential and becomes strongly repulsive for nearby points at the Fermi surface. This leads to an overall gap suppression despite the increased interaction strength.

\begin{figure}[t]
\centering
 	\includegraphics[angle=0,width=\columnwidth]{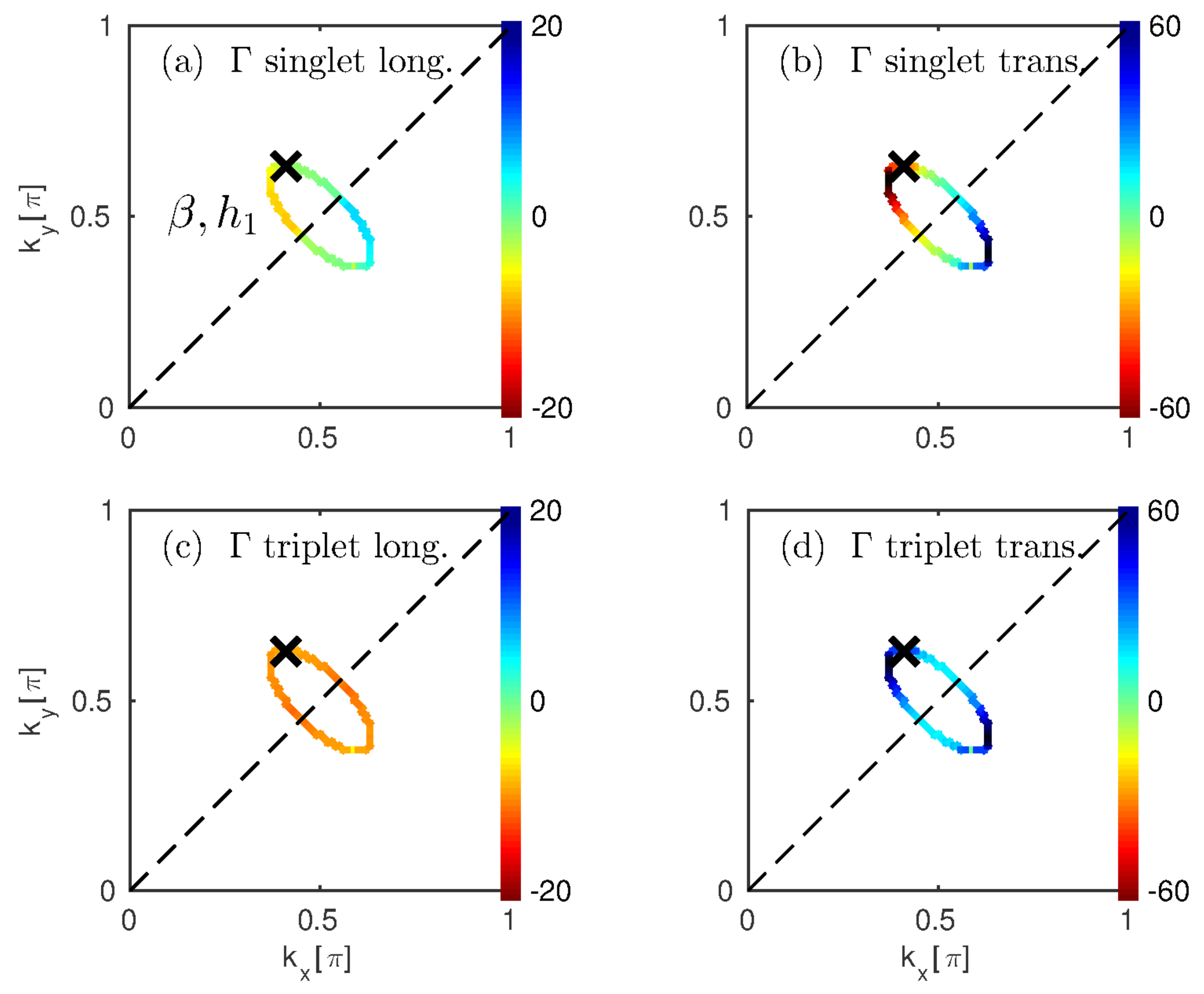}
 	\caption{(color online) Effective pairing contributions from longitudinal (a,c) and transverse (b,d) spin fluctuations in the singlet and triplet channel on the hole pocket centered at $(\frac{\pi}{2},\frac{\pi}{2})$. The black cross denotes the position of $\kv'$ and the value of $\Gamma_{\kv,\kv'}^{s/t}$ is shown as a function of $\kv$ around the hole pocket. Negative potential contributions correspond to effective attraction. The zone diagonal is indicated by a dashed line. The filling is $\Nav=0.96$ and $U=3.0$, $t'=-0.35$. Note the different color scale for pairing potentials in the longitudinal and transverse channels.}
\label{fig:potBeta}
\end{figure}
\begin{figure}[b]
\centering
 	\includegraphics[angle=0,width=\columnwidth]{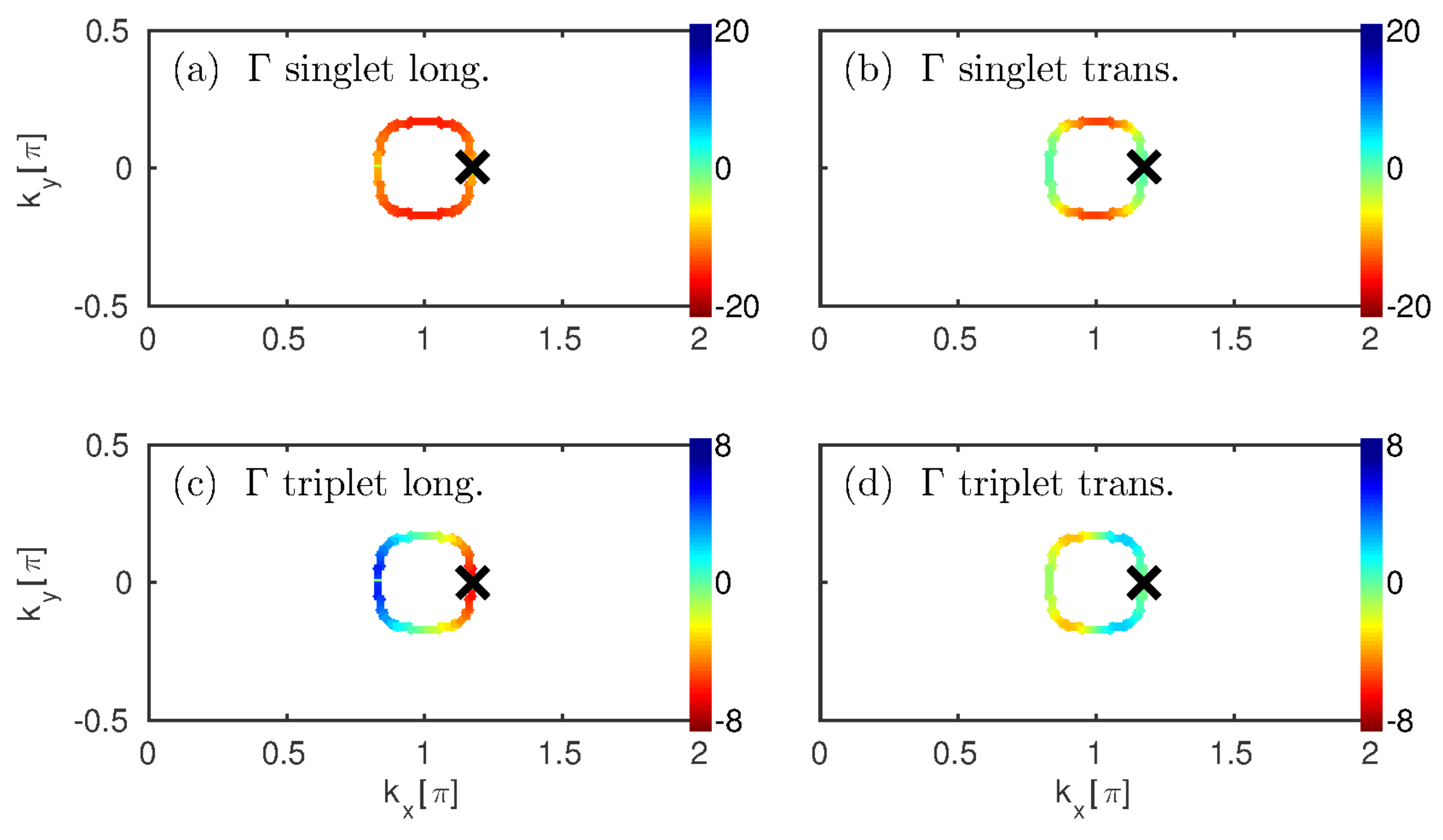}
 	\caption{(color online) Effective pairing contributions from longitudinal (a,c) and transverse (b,d) spin fluctuations in the singlet and triplet channel on the electron pocket centered at $(\pi,0)$. The black cross denotes the position of $\kv'$ and the value of $\Gamma_{\kv,\kv'}^{s/t}$ is shown as a function of $\kv$ around the electron pocket. The filling is $\Nav=1.05$ and $U=3.0$, $t'=-0.35$. Note the different color scale for singlet and triplet pairing potentials.}
\label{fig:potAlpha}
\end{figure}

In general, knowledge of the $(\kv,\kv')$ structure of the pairing potential is crucial in order to decide what kind of superconducting instabilities are favored in the AF phase. To address this question,  in Figs. \ref{fig:potBeta} and  \ref{fig:potAlpha} we map out both the singlet and triplet pair scattering potentials due to longitudinal and transverse spin fluctuations for hole- and electron doping, respectively. In the case of small hole doping (Fig. \ref{fig:potBeta}), the singlet pairing interactions from longitudinal and transverse spin fluctuations are similar in structure; there is an effective attractive (negative) interaction for $\kv$ and $\kv'$ on the same side of the zone diagonal, whereas the interaction becomes repulsive (positive) when $\kv$ and $\kv'$ are on opposite sides of the zone diagonal, as deduced from Fig.~\ref{fig:potBeta}(a-b). This structure supports $d$-wave as well as $g$-wave states.
In this case, the pairing interaction due to transverse spin fluctuations is much stronger than the pairing arising from longitudinal spin fluctuations, as was also deduced from Fig.~\ref{fig:potphasediagram}, and expected from Refs.~\onlinecite{WenyaNJP},~\onlinecite{Frenkel90} and~\onlinecite{ChubukovFrenkel}. In the triplet channel, the effective interaction arising from the two different types of spin fluctuations has very different character.
The longitudinal fluctuations in fact give rise to a locally attractive pairing on the entire pocket, as shown in Fig.~\ref{fig:potBeta}(c). This type of pairing supports a nodeless gap of $p$-wave character, i.e. $\Delta_{\kv} \propto \sin(k_x)$. However, due to the strong intrapocket repulsive interaction  mediated by transverse spin fluctuations shown in Fig.~\ref{fig:potBeta}(d),   the total pairing interaction does not allow for a $p$-wave solution.~\cite{WenyaNJP} Therefore the triplet solution becomes higher order and will display nodes at the Fermi level.
\begin{figure}[b]
\centering
 	\includegraphics[angle=0,width=\columnwidth]{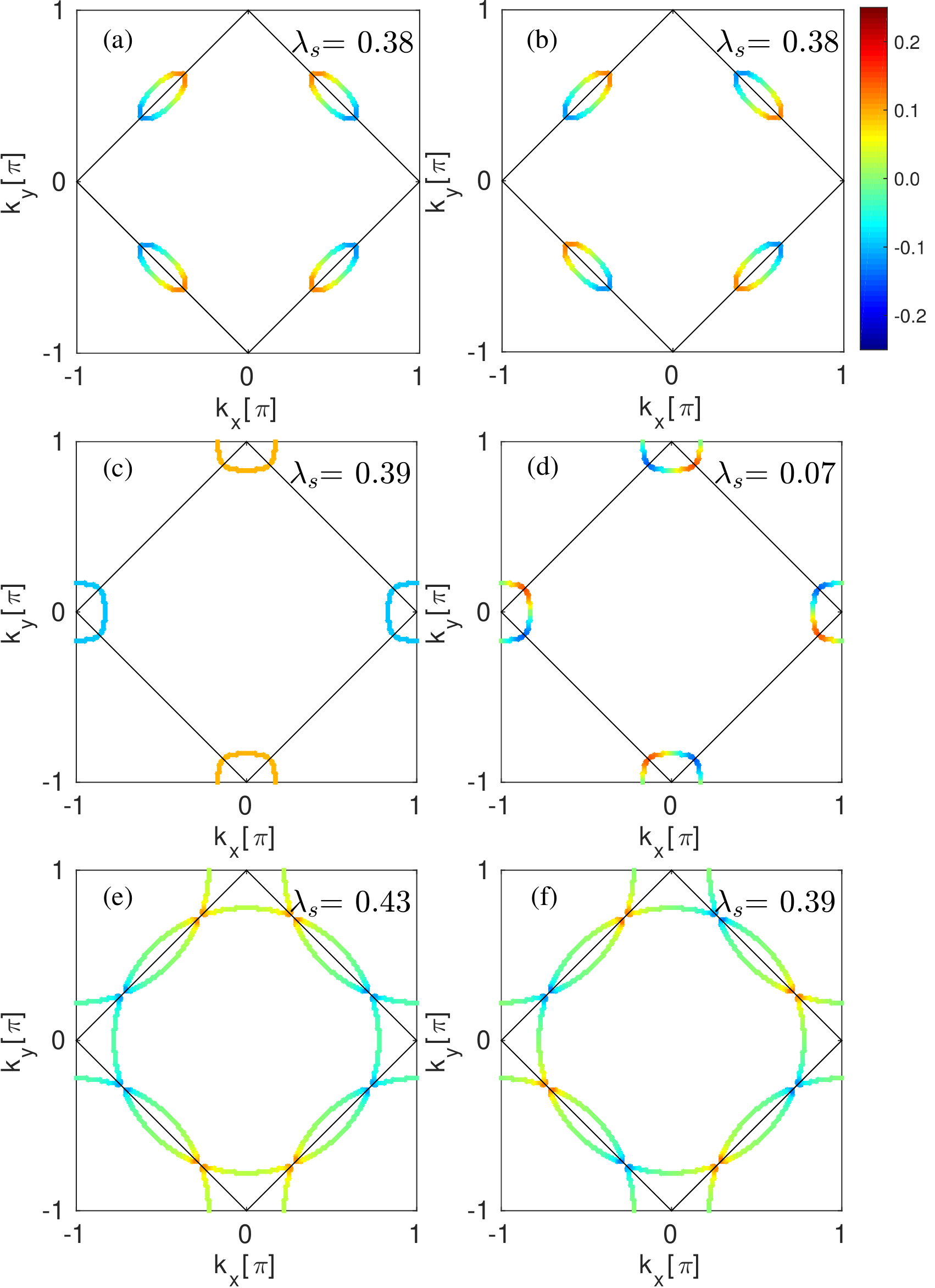}
\caption{(color online) Solutions to the linearized gap equation in the presence of (a,b) only hole pockets (4 \% hole doping and $U=3$), (c,d) only electron pockets (5 \% electron doping and $U=3$) or (e,f) both types of pockets (5\% electron doping and $U=2.735$).
 	The last situation is obtained very close to the AF quantum critical point, i.e. for $W \to 0$. The leading solutions are shown in the first column and the sub-leading solutions are shown in the second column. In all three cases the leading solution is $d_{x^2-y^2}$, and the sub-leading solution is $g$-wave. The latter differs from $s$-wave by being odd under $k_x \to -k_x$  and  $k_y \to -k_y$.
 	Note that in the case of hole doping the two solutions $d_{x^2-y^2}$ and $g$ are nearly degenerate with $\lambda_{d}=0.384$ and $\lambda_{g}=0.381$. In the case of small and intermediate electron doping, the $d_{x^2-y^2}$ solution becomes strongly dominant. Upon increased electron doping approaching the quantum critical point, the $g$-wave solution becomes increasingly important, although always subdominant.}
\label{fig:solutionsLGE}
\end{figure}

For electron doping, the pairing interaction in the singlet channel is locally attractive for $\kv$ and $\kv'$ residing on the same electron pocket, as shown in Fig.~\ref{fig:potAlpha}(a-b). Attractive pairing occurs due to both longitudinal and transverse spin fluctuations. As opposed to the hole-doped case, it is the longitudinal fluctuations that mediate the strongest effective pairing. In this regime, it is clearly the $d_{x^2-y^2}$ solution which will be favored due to the symmetry $\Gamma_{\kv,\kv'+\Qv}=-\Gamma_{\kv,\kv'}$. Upon increasing the electron doping the purely attractive intrapocket interaction is replaced by partly repulsive regions and this will diminish the resulting superconducting gap value.
In the triplet channel the pairings from both types of fluctuations will be quite weak, see Fig.~\ref{fig:potAlpha}(c-d). As in the case of hole doping, the longitudinal fluctuations support a $p$-wave gap since it is attractive for $\kv$ and $\kv'$ located at the same side of the Fermi pocket and repulsive for $\kv$ and $\kv'$  on opposite sides, as shown in Fig.~\ref{fig:potAlpha}(c). The transverse fluctuations display the reverse structure as evident from in Fig.~\ref{fig:potAlpha}(d), which is the reason why the $p$-wave solution becomes suppressed also on the electron-doped side.
\subsection{Solutions to the linearized gap equation}
\label{sec:lgeresults}
Keeping in mind the structure of the pairing interactions described above, we now turn to the solutions to the linearized gap equation Eq.~(\ref{eq:lge}) in order to determine the subleading solutions and the doping evolution of the three leading solutions. We consider three qualitatively different types of Fermi surfaces. First, we study a Fermi surface consisting of hole pockets at $(\pm \frac{\pi}{2},\pm \frac{\pi}{2})$ which occurs on the hole-doped side, see Fig.~\ref{fig:solutionsLGE}(a). Second, we turn to electron pockets at $(\pm \pi,0)$ and $(0,\pm \pi)$ which occur for small and intermediate electron doping levels, see Fig.~\ref{fig:solutionsLGE}(c), and third, we study the occurrence of both electron and hole pockets very close to critical doping as in Fig.~\ref{fig:solutionsLGE}(e), i.e. the doping level for which AF order disappears. As discussed above, the presence of hole pockets destabilizes the commensurate AF by additional intrapocket contributions to the bare spin susceptibilities. In this case, we force stability of the commensurate AF order by turning off the intrapocket contributions by hand, which is justified from the work by Chubukov and Frenkel\cite{ChubukovFrenkel} showing that vacuum renormalization leads to a stability of the commensurate AF order at small hole dopings.

By inspection of the intrapocket potential structure on the hole pocket shown in Fig.~\ref{fig:potBeta}(a,b),
we expect either of the two solutions $d_{x^2-y^2}$ or $g$ for the weakly hole-doped system. From the intrapocket potential structure it is not possible to qualitatively distinguish between these two solutions. The only other way that the system might choose one solution over the other would be from the structure of the pairing potential for $\kv$ on one hole pocket and $\kv'$ on the neighboring pocket. However, from the numerical potential evaluation it turns out that the interpocket pairing contribution is an order of magnitude smaller than the intrapocket pairing and its symmetry does not favor one of the solutions over the other. This is the reason we obtain two nearly degenerate singlet solutions, namely $d_{x^2-y^2}$ and the $g$ solutions, in Fig.~\ref{fig:solutionsLGE}(a,b).

Turning to the case of 5 \% electron doping shown in Fig.~\ref{fig:solutionsLGE}(c,d), we see that the $d_{x^2-y^2}$ solution is clearly favored over the sub-leading solution $g$, which becomes strongly suppressed. This is a direct consequence of the intrapocket attraction between $\kv$ and $\kv'$ residing on the same electron pocket.
Upon increasing doping, the electron pockets grow in size, but throughout the SDW region, the $d_{x^2-y^2}$ solution continues to be the leading solution although the sub-leading $g$-wave gets closer. Also close to the crossover to the paramagnetic phase, where both electron and hole pockets are present at the Fermi surface,
the leading solution remains $d_{x^2-y^2}$ as shown in Fig.~\ref{fig:solutionsLGE}(e,f).

\begin{figure}[b]
\centering
 	\includegraphics[angle=0,width=0.875\linewidth]{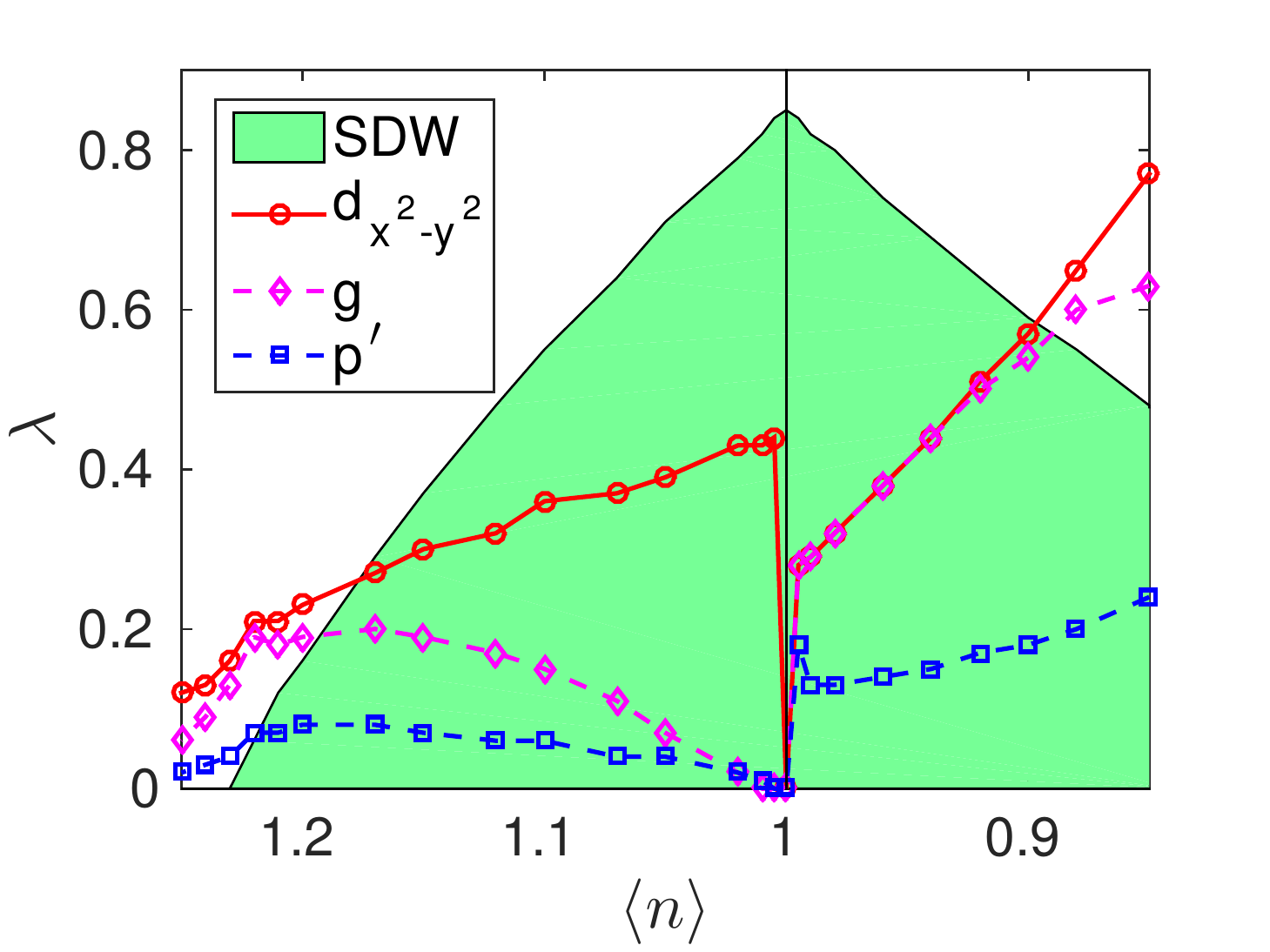}
 	\caption{(color online) Three leading superconducting instabilities as a function of filling. The SDW region is shown by the green area. The three largest eigenvalues to the linearized gap equation, Eq.~(\ref{eq:lge}), are shown. The $d_{x^2-y^2}$ solution (red line) dominates at all fillings. The next-nearest neighbor hopping is $t'=-0.35$ and the bare Coulomb interaction is $U=3$. }
 	\label{fig:phasediagramU3nt0p35}
 	\end{figure}

\subsection{Coexistence phase diagram}
The complete doping evolution of the three leading superconducting order parameters, $d_{x^2-y^2}$, $g$, and $p'$-wave, is shown in the phase diagram in Fig.~\ref{fig:phasediagramU3nt0p35}. On the hole-doped side, we limit the study to the underdoped region. 
Very close to half filling we observe a near-degeneracy of the $d_{x^2-y^2}$ and $g$-wave solutions.
The existence of two degenerate solutions in principle
allows for their mixture as a TRSB solution, or for orthorhombic distortions as a direct mixture. At the mean field level, the preference of a low temperature TRSB solution might be expected when it is possible to remove gap nodes at the Fermi surface, since removal of gaps from $|\Delta_\kv|$ leads to a gain in the condensation energy.
In the present case of the degenerate solutions $d_{x^2-y^2}$ and $g$, however,  we note that these share the same nodes along the zone diagonal. Thus, there will be no apparent gain of energy by constructing a TRSB solution of the form $d_{x^2-y^2} \pm i g$. In fact, solving the full gap equation as stated in Eq.~(\ref{eq:SCGapEquation}) reveals that the
$d_{x^2-y^2}$ solution is energetically favored.
Upon larger hole doping, the near-degeneracy of the $d_{x^2-y^2}$ and $g$-wave solutions is split and the  $d_{x^2-y^2}$ solution becomes clearly dominant in this regime.
In the case of electron doping, the $d$-wave solution is strongest very close to half filling even though this is not where the longitudinal and transverse pairing potentials achieve their maximum strengths, see Fig.~\ref{fig:potphasediagram}. The reason is that in the limit where electron pockets are small, the structure of the intrapocket pairing potentials is purely attractive, as shown in Fig.~\ref{fig:potAlpha}(a,b), and this strongly supports a $d$-wave solution.
We note that this feature of a well-developed gap in the limit of small electron dopings is an inherent result of the weak-coupling approach to the coexistence phase.
At critical electron doping for which $W\to 0$, the Fermi arcs just touch the magnetic zone boundary and as a consequence,
nesting by $\Qv$ on the paramagnetic side is rapidly weakened upon increased electron doping. The ordering of the leading solutions remains the same as in the SDW phase, with the $p'$ solution, which in the paramagnetic phase takes the simpler form $[\cos(k_x)-\cos(k_y)]\sin(k_x)$, the least favorable. The paramagnetic Fermi surface, which is a hole pocket centered at $(\pi,\pi)$ is roughly circular thereby preventing nesting not only at $\Qv$, but at any $\qv$-vector.
As a result spin-fluctuation-mediated superconductivity rapidly dies off.
The evolution of the $d_{x^2-y^2}$ solution was previously discussed in Ref.~\onlinecite{Khodel04}, where the possibility of a different pairing mechanism close to the crossover between SDW and paramagnetism was speculated. In Ref.~\onlinecite{Khodel04} the pairing was treated at a phenomenological level and here we note that also the full treatment of the spin-mediated pairing gives rise to a rapid decrease in the superconductivity upon entering the paramagnetic region.

\begin{figure}[t!]
\centering
 	\includegraphics[angle=0,width=0.875\linewidth]{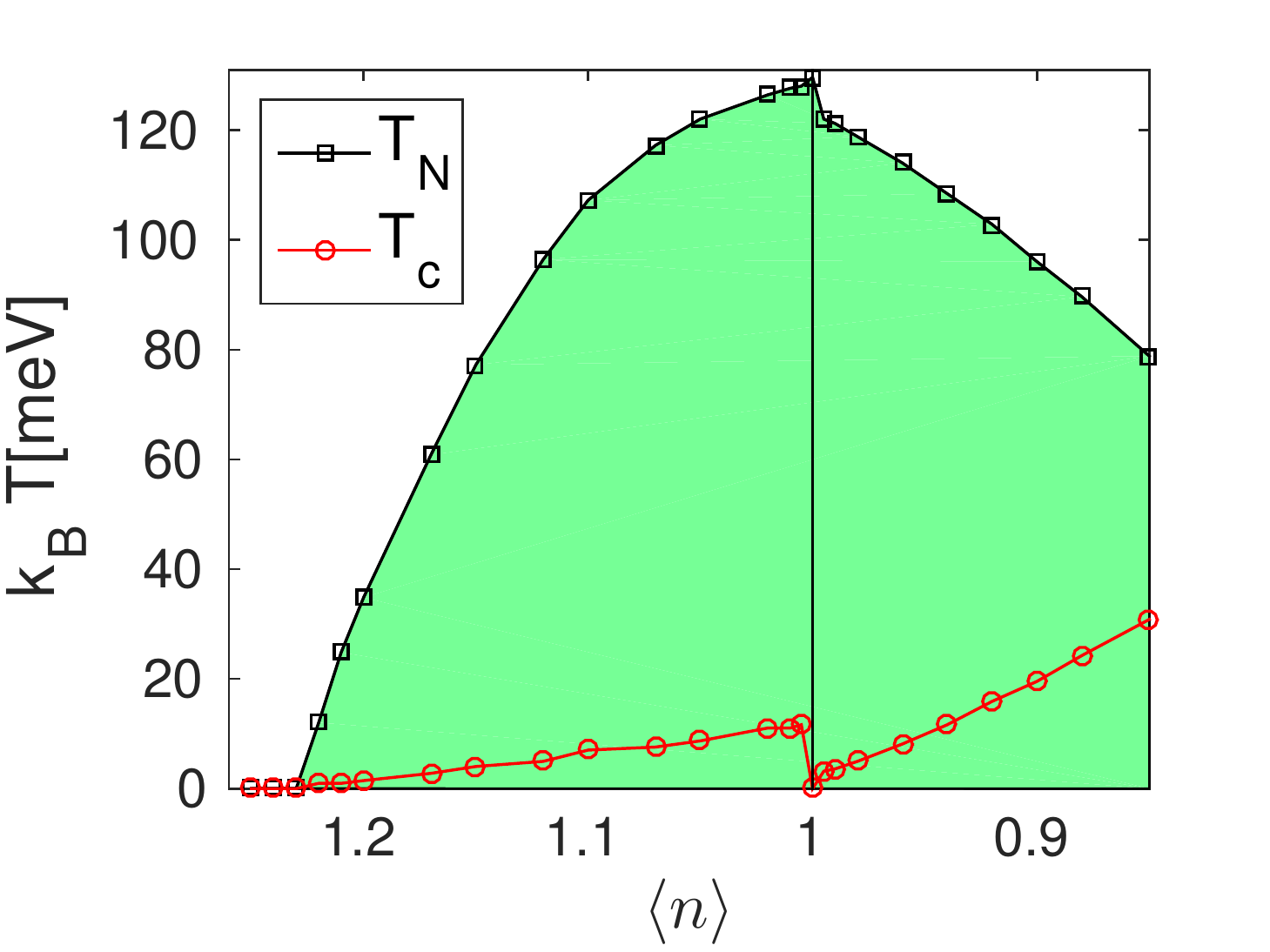}
 	\caption{(color online) The critical temperature, $T_c=1.13 \epsilon_c e^{-1/\lambda}$, with the energy cut off set to $\epsilon_c=0.25$ and $t=400$ meV. The superconducting instability is  $d_{x^2-y^2}$ at all fillings. The SDW region which is determined from the mean field gap equation is shown by the green area. The next-nearest neighbor hopping is $t'=-0.35$ and the bare Coulomb interaction is $U=3$. Note that within the weak-coupling approach the SDW ground state is always metallic away from $\Nav=1$ and therefore, does not cover the Mott insulating behavior of in the underdoped region of hole-doped cuprates.}
 	\label{fig:TempphasediagramU3nt0p35}
 	\end{figure}

Finally, we show estimations for the critical temperatures of the commensurate AF order, $T_N$, as well as the superconducting ordering temperature, $T_c$ in Fig.~\ref{fig:TempphasediagramU3nt0p35}. The $T_c$ is estimated from the eigenvalue $\lambda$ obtained for the leading $d$-wave solution as a function of doping. Whereas $T_c$ drops off upon increasing electron doping it shows the opposite evolution as a function of hole doping. At $\Nav =1$ the superconducting instability is absent due to a full gapping of the Fermi surface by the magnetic order. Similar behavior was found by variational cluster perturbation theory in the work by S\'en\'echal and coworkers.~\cite{Senegal}

As discussed in the Introduction, the results on the hole-doped side contradicts the recent findings of Ref.~\onlinecite{Lu2014}, where a $p$-wave solution appears as the leading instability. In fact, the Cooper-pairing in Ref.~\onlinecite{Lu2014} is mediated by an exchange interaction which involves only nearest neighbor sites, while in our approach
the interaction potential in the Cooper channel invokes effectively also sites which are farther apart.
The similarity of the treatment of the coexistence phase in Ref.~\onlinecite{Lu2014} and
our approach allows for a direct comparison of the interaction Hamiltonians. In the simplified case of only hole pockets at the Fermi surface, the interaction Hamiltonian reads generally
\begin{eqnarray}
  H_{\rm hole}&=&\sum_{\kv,\kv'\spin} \Gamma(\kv,\kv') \beta_{\kv\spin}^\dagger \beta_{-\kv\ospin}^\dagger\beta_{-\kv'\ospin}\beta_{\kv'\spin}.\label{generalHam}
\end{eqnarray}
In the {\it t-J}-like model without double occupancy constraint employed in Ref.~\onlinecite{Lu2014}, the effective interaction entering Eq.~\eqref{generalHam} takes the form
\begin{eqnarray}
\Gamma_{\rm t-J}(\kv,\kv') &=&-\frac{J(\kv-\kv')}{2}
[m^2(\kv,\kv')+l^2(\kv,\kv')] \nonumber \\
&&-J(\kv+\kv')[n^2(\kv,\kv')+p^2(\kv,\kv')], \nonumber \\
\label{eq:LuGamma}
\end{eqnarray}
with $J(\qv)=J[\cos(q_x)+\cos(q_y]$ and $J>0$. By contrast, in the Hubbard model the effective interaction entering Eq.~\eqref{generalHam} is given by
\begin{eqnarray}
\Gamma_{\rm Hub}(\kv,\kv') &=& \Gamma^z_{\kv,\kv'}\pm 2 \Gamma^{+-}_{\kv,\kv'},\label{eq:betaHub}
\end{eqnarray}
with the longitudinal and transverse effective interactions stated in Eqs.~(\ref{eq:GammaZ}) and (\ref{eq:GammaPM}). The lower sign of Eq.~(\ref{eq:betaHub}) belongs to the triplet channel and this is the source of an effective intrapocket repulsion on the hole-doped side as shown in Fig.~\ref{fig:potBeta}(d). On the contrary, the effective interaction as stated in Eq.~(\ref{eq:LuGamma}) gives rise to a purely attractive triplet potential for $\kv$ and $\kv'$ residing on the same hole pocket, and therefore the $p$-wave solution would indeed appear
to be the dominating instability. 
Note, however, that the model of Ref.~\onlinecite{Lu2014} cannot be considered strictly as a strong-coupling limit of the single band Hubbard model as it does not include a constraint for no double occupancies of the fermions explicitly. This could play an important role. For example, as mentioned in the Introduction, the original strong-coupling study of unconventional superconductivity driven by the spin waves, studied within the {\it t-J} model with the constraint of no double occupancies, does find the d$_{x^2-y^2}$-wave symmetry of the superconducting gap to be the only stable solution.~\cite{Sushkov,Sushkov2}

\subsection{Anomalous pairing gap}
\label{sec:anomalouspairingresults}
We introduced the possibility of additional pairing gaps in the coexistence phase in Sec.~\ref{sec:anomalouspairing}. Such anomalous pairings were neglected in the previous discussion, see Figs.~\ref{fig:potphasediagram}-\ref{fig:TempphasediagramU3nt0p35}, where we demonstrated that the dominating symmetry of the normal intraband pairing gaps is $d_{x^2-y^2}$ at all doping values.
Now we address the question of whether the inclusion of anomalous interband gaps introduces essential modifications to this result.
The anomalous gaps develop under the constraint that they must have the same parity as the normal intraband gaps. In the current case, this means that anomalous pairing gaps must necessarily be of even parity.

\begin{figure}[t!]
\centering
 	\includegraphics[angle=0,width=0.875\linewidth]{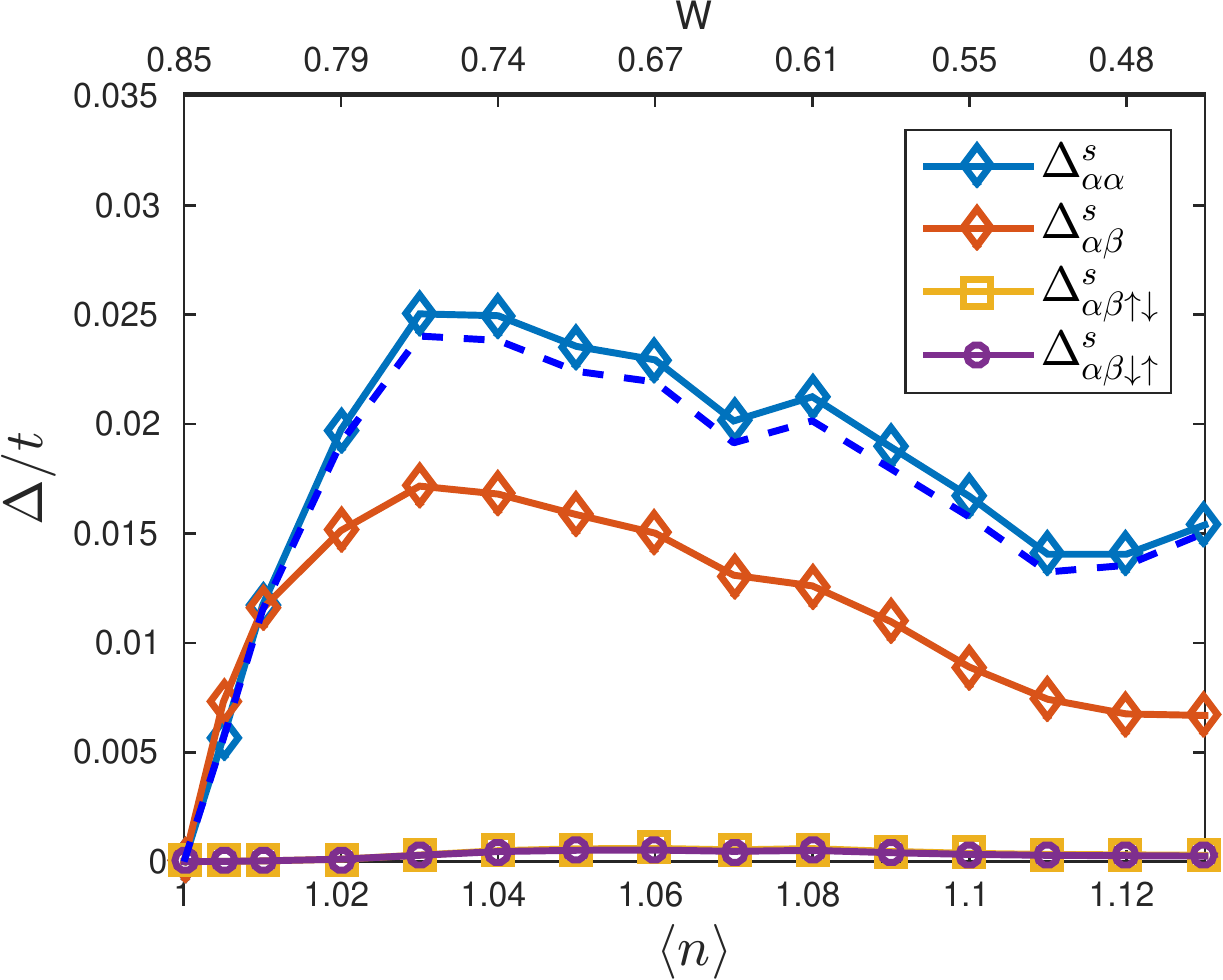}
 	\caption{(color online) Self-consistent solution for the normal and anomalous superconducting gaps as stated in Eqs.~(\ref{eq:Daa})-(\ref{eq:gapanodownup}) for the electron-doped system with $\beta$-bands outside the energy cut off. The superconducting gap symmetry is $d_{x^2-y^2}$ for both normal and anomalous gaps. The next-nearest neighbor hopping is $t'=-0.35$ and the bare Coulomb interaction is $U=3$. The energy cut off set to $\epsilon_c=0.15$. We show the absolute value of the gap averaged over $\kv$-states within the energy range $\pm \epsilon_c$, e.g. $\Delta_{\alpha \alpha}^s=\frac{1}{N_s} \sum_\kv |\Delta_{\alpha \alpha}^s (\kv)|$. Within this energy range no normal superconducting gap develops on the $\beta$-band, i.e. $\Delta_{\beta\beta}^s=0$. The dashed blue line shows $\Delta_{\alpha \alpha}^s$ in the case where all anomalous gaps are excluded.}
 	\label{fig:ElectronSelfconsistent}
\end{figure}

We solve the system of five coupled gap equations as stated in Eqs.~(\ref{eq:Daa})-(\ref{eq:gapanodownup}) numerically to obtain the fully self-consistent solutions in the coexistence phase. Since we have not included any energy dependence of the effective pairing vertices, we restrict the momentum sums of Eqs.~(\ref{eq:Daa})-(\ref{eq:gapanodownup}) to states in the vicinity of the Fermi surface. The effective potential for pair scattering between anomalous gaps, $ \Gamma^{\alpha\beta\alpha\beta,2}_{\kv,\kv'}$ stated in Eq.~(\ref{eq:GammaABAB2}), formally includes a divergent contribution for $\kv'=-\kv$. To avoid such contributions we cut off all pairings $ \Gamma^{\alpha\beta\alpha\beta,2}_{\kv,\kv'}$ to a maximum value of $V_{\rm max}=200t$.
By this procedure, we find that also the anomalous gaps acquire a  $d_{x^2-y^2}$ structure.
In Fig.~\ref{fig:ElectronSelfconsistent} we present the normal and anomalous mean field gaps given in Eqs.~(\ref{eq:Daa})-(\ref{eq:gapanodownup}) in the case of an electron-doped system in the regime of well-developed SDW order. We show the gap value averaged over $\kv$-states close to the Fermi surface, e.g. $\Delta_{\alpha \alpha}^s=\frac{1}{N_s} \sum_\kv |\Delta_{\alpha \alpha}^s (\kv)|$, where $N_s$ is the number of $\kv$-states with $|E^\alpha_\kv|<\epsilon_c$ and $\epsilon_c=0.15t$ is the cut off energy. Upon approaching half filling where $W=0.85$ for this band, the self-consistent solutions gradually decrease. This differs from the results obtained for the linearized gap equation in Fig.~\ref{fig:phasediagramU3nt0p35} where the decrease is very abrupt. The gradual decrease in Fig.~\ref{fig:ElectronSelfconsistent} follows the decrease in the number of states within the energy range around the Fermi surface towards half filling.
\begin{figure}[t!]
\centering
 	\includegraphics[angle=0,width=0.95\linewidth]{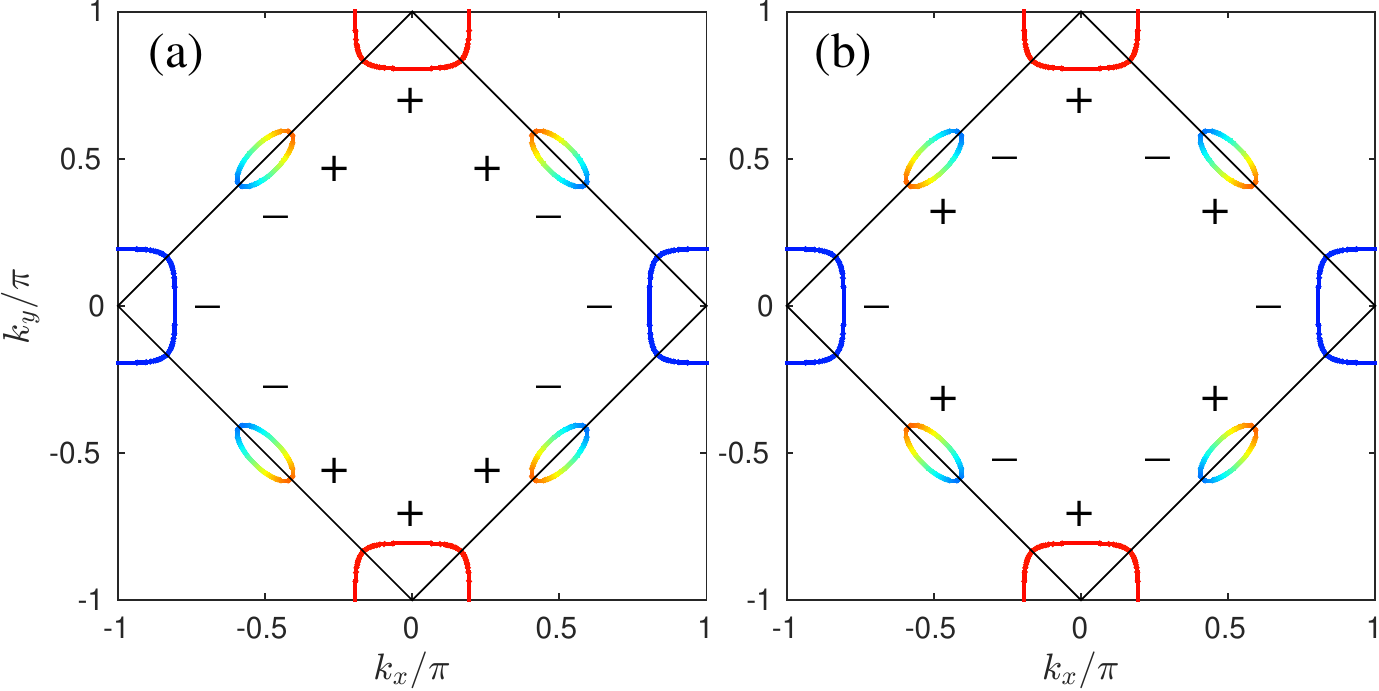}
 	\caption{(color online) Form factors of (a) the normal superconducting gaps $\Delta_s^{\alpha \alpha}$ and $\Delta_s^{\beta \beta}$ and (b) the anomalous gap, $\Delta_s^{\alpha \beta}$. All gaps have $d_{x^2-y^2}$ structure. Whereas the normal gaps show the same overall phase on electron and hole pockets, the anomalous gaps display an internal $\pi$-shift between the electron and hole pockets.}
 	\label{fig:Phase}
\end{figure}

From the results presented in Fig.~\ref{fig:ElectronSelfconsistent}, it is evident that the mean field $\Delta_{\alpha \beta}^s$ is the most important contributor to anomalous pairing whereas $\Delta_{\alpha \beta \up \down}^s$ and $\Delta_{\alpha \beta \down \up}^s$ are negligible. This illustrates the point that it is the linear coupling to the normal gaps via the pairing interaction $\Gamma^{\alpha \beta}_{\kv,\kv'}$ of Eq.~(\ref{eq:pairingAAAB}) which generates the anomalous pairings. Without this coupling term also the subdominant mean fields $\Delta_{\alpha \beta \up \down}^s$ and $\Delta_{\alpha \beta \down \up}^s$ would vanish. Furthermore, the inclusion of anomalous pairings slightly enhances the size of the intraband gap. This is seen from Fig.~\ref{fig:ElectronSelfconsistent} by compairing $\Delta_{\alpha \alpha}^s$ as deduced from the self-consistent determination of all five gaps shown by blue circles with a full line with the self-consistent calculation of the normal intraband gaps alone, which is shown by the blue dashed line. In the latter case the gap magnitudes are slightly smaller. 

If hole pockets are included within the energy cut off, i.e. $|E_\kv^\beta|<\epsilon_c$, the self-consistent calculation finds a $\pi$-shift of the anomalous superconducting phase at the hole pockets compared to the electron pockets, see Fig.~\ref{fig:Phase} (b). This is a robust feature arising from the momentum structure of the normal-anomalous pair scattering potential and as a result of the internal sign between $\alpha$- and $\beta$-operators in Eq.~(\ref{eq:gapano}). We have tested numerically that this result is insensitive to the relative ratio of normal and anomalous pairing potentials. 

Lastly, we mention that the self-consistent gap solutions are all real-valued. Therefore unlike the iron-based case studied in Ref.~\onlinecite{Hinojosa}, we do not find a TRSB superconductor in the coexistence phase. In conclusion, the incorporation of anomalous pairings in the effective interaction Hamiltonian gives only small quantitative changes to the intraband gap results presented in the previous sections.

\section{CONCLUSIONS}
This study represents a very detailed investigation of spin-fluctuation-mediated superconductivity in a system with well-developed itinerant antiferromagnetic order and 
includes a calculation of the complete coexistence phase diagram of
the Hubbard model. The spin fluctuations bear important fingerprints of the spin order with the transverse spin fluctuations corresponding to the Goldstone mode of the spin-symmetry-broken state. 
We find that longitudinal and transverse spin fluctuations are equally important for the development of an effective pairing glue between the fermionic quasi-particles of the spin-density-ordered metal. Both types of spin fluctuations act in concert in the even parity quasi-spin singlet channel and this gives rise to a robust gap solution of $d_{x^2-y^2}$ structure at all doping levels within the coexistence doping region. The situation is quite different in the case of odd parity quasi-spin triplet superconductivity. Here longitudinal spin fluctuations promote a nodeless $p$-wave solution in the hole-doped system due to an effective intrapocket attraction. However, a strong intrapocket repulsive contribution from transverse fluctuations destroys this effective attraction and destabilizes the $p$-wave solution. Thus, the coexistence phase treated within the Hubbard model does not support a nodeless gap on the hole doped side, in contrast to a recent study of the coexistence phase within the $t-J$ model without double occupancy constraint.~\cite{Lu2014}
A modified version of the Hubbard model in which the relative contribution from the transverse fluctuations is suppressed, could also stabilize a nodeless $p$-wave state.

Finally, we have investigated the additional interband pairing amplitudes that appear in the SDW phase as a consequence of the linear coupling to the normal intraband gaps. Such anomalous pairing gaps are required  to have the same parity under inversion as the normal intraband gaps. Specifically, the anomalous pairing
amplitudes also manifest a $d_{x^2-y^2}$ structure and the gap magnitude is similar to the normal pairing gaps.

\section*{Acknowledgements}

We thank A.V. Chubukov, A. Kreisel, and S. Mukherjee for useful discussions, and W. Rowe for her contributions to the early stages
of this project.
A. T. R. and B. M. A acknowledge support from a Lundbeckfond fellowship (Grant A9318).
P. J. H. acknowledges support from NSF-DMR-1005625.
The work of I. E. was supported by the Focus Program 1458 Eisen-Pniktide of the DFG, and by the German Academic
Exchange Service (PPP USA no. 57051534). I. E. acknowledges the financial support of the Ministry of
Education and Science of the Russian Federation in the framework of
Increase Competitiveness Program of NUST MISiS (No. 2-2014-015)


\begin{thebibliography}{10}

\bibitem{IsmerPRL10} J.-P. Ismer, I. Eremin, E. Rossi, D. K. Morr, and G. Blumberg, Phys. Rev. Lett. {\bf 105}, 037003 (2010).

\bibitem{Schmiedt14} J. Schmiedt, P. M. R. Brydon, and C. Timm, Phys. Rev. B {\bf 89}, 054515 (2014).

\bibitem{WenyaNJP} W. Rowe, I. Eremin, A. T. R\o mer, B. M. Andersen, and P. J. Hirschfeld, New J. Phys. {\bf 17}, 023022 (2015).

\bibitem{Schrieffer89} J. R. Schrieffer, X. G. Wen, and S. C. Zhang,  Phys. Rev. B {\bf 39}, 11663 (1989).

\bibitem{Frenkel90}  D. M. Frenkel and W. Hanke, Phys. Rev. B {\bf 42}, 6711 (1990).

\bibitem{review-greene} N. P. Armitage, P. Fournier, and R. L. Greene, Rev. Mod. Phys. {\bf 82}, 2421 (2010).

\bibitem{Manske00} D. Manske, I. Eremin, and K. H. Bennemann
Phys. Rev. B {\bf 62}, 13922 (2000).

\bibitem{Khodel04} V. A. Khodel, V. M. Yakovenko, M. V. Zverev, and H. Kang, Phys. Rev. B {\bf 69}, 144501 (2004).

\bibitem{Yoshimura} H. Yoshimura and D. S. Hirashima, Jour. Phys. Soc. Japan {\bf 73}, 2057 (2004).

\bibitem{Parker08} D. Parker and A. V. Balatsky, Phys. Rev. B {\bf 78}, 214502 (2008).

\bibitem{Kyung03} B. Kyung, J.-S. Landry, and A.-M. S. Tremblay, Phys. Rev. B {\bf 68}, 174502 (2003).

\bibitem{Guinea} F. Guinea, R. S. Markiewicz, and M. A. H. Vozmediano, Phys. Rev. B {\bf 69}, 054509 (2004).

\bibitem{Astrid2015} A. T. R\o mer, A. Kreisel, I. Eremin, M.A. Malakhov, T.A. Maier, P.J. Hirschfeld, and B.M. Andersen, Phys. Rev. B {\bf 92}, 104505 (2015).

\bibitem{Matsui05} H. Matsui, K. Terashima, T. Sato, T. Takahashi, M. Fujita, and K. Yamada, Phys. Rev. Lett. {\bf 95}, 017003 (2005).

\bibitem{Blumberg02} G. Blumberg, A. Koitzsch, A. Gozar, B. S. Dennis, C. A. Kendziora, P. Fournier, and R. L. Greene, Phys. Rev. Lett. {\bf 88}, 107002 (2002).

\bibitem{Krotkov} P. Krotkov and A. V. Chubukov, Phys. Rev. Lett. {\bf 96}, 107002 (2006).

\bibitem{Das06} T. Das, R. S. Markiewicz, and A. Bansil, Phys. Rev. B {\bf 74}, 020506(R) (2006).

\bibitem{Ting06} Q. Yuan, F. Yuan, and C. S. Ting, Phys. Rev. B {\bf 73}, 054501 (2006).

\bibitem{ChubukovFrenkel}  A. V. Chubukov and  D. M. Frenkel  Phys. Rev. B {\bf 46}, 11884 (1992).

\bibitem{RowePRB12} W. Rowe, J. Knolle, I. Eremin, and P. J. Hirschfeld, Phys. Rev. B {\bf 86}, 134513 (2012).

\bibitem{Scalapino86} D. J. Scalapino, E. Loh, Jr., and J. E. Hirsch, Phys. Rev. B {\bf 34} 8190 (1986).

\bibitem{berkschrie}  N. F. Berk and J. R. Schrieffer, Phys. Rev. Lett. {\bf 17}, 433 (1966).

\bibitem{Hlubina} R. Hlubina, Phys. Rev. B {\bf 59}, 9600 (1999).

\bibitem{Chubukov} A. V. Chubukov and J.-P. Lu, Phys. Rev. B {\bf 46}, 11163 (1992).

\bibitem{Markiewicz08} R. S. Markiewicz and A. Bansil, Phys. Rev. B {\bf 78}, 134513 (2008).

\bibitem{Raghu10} S. Raghu, S. A. Kivelson, and D. J. Scalapino, Phys. Rev. B {\bf 81}, 224505 (2010).

\bibitem{Eberlein14} A. Eberlein and W. Metzner, Phys. Rev. B {\bf 89}, 035126 (2014).

\bibitem{Luschersushkov} A. L\"uscher, A. I. Milstein, and O. P. Sushkov, Phys. Rev. B {\bf 75}, 235120 (2007).

\bibitem{Choetal} W. Cho, R. Thomale, S. Raghu, and S. A. Kivelson Phys. Rev. B {\bf 88}, 064505 (2013).

\bibitem{Galitski} V. Galitski and S. Sachdev, Phys. Rev. B {\bf 79}, 134512 (2009).

\bibitem{Lu2014} Y.-M. Lu, T. Xiang, and D.-H. Lee, Nature Phys. {\bf 10}, 634 (2014).

\bibitem{razzoli} E. Razzoli, G. Drachuck, A. Keren, M. Radovic, N. C. Plumb, J. Chang, Y.-B. Huang, H. Ding, J. Mesot, and M. Shi, Phys. Rev. Lett. {\bf 110}, 047004 (2013).

\bibitem{peng} Y. Peng, J. Meng, D. Mou, J. He, L. Zhao, Y. Wu, G. Liu, X. Dong, S. He, J. Zhang, X. Wang, Q. Peng, Z. Wang, S. Zhang, F. Yang, C. Chen, Z. Xu, T. K. Lee, and X. J. Zhou, Nat. Comms. {\bf 4}, 2459 (2013).

\bibitem{Sushkov} M. Yu. Kuchiev and O.P. Sushkov, Physica C {\bf 218}, 197 (1993); V.V. Flambaum, M.Yu. Kuchiev, and O.P. Sushkov,
Physica C {\bf 227}, 267 (1994); V. I. Belinicher, A. L. Chernyshev, A. V. Dotsenko, and O. P. Sushkov, Phys. Rev. B {\bf 51}, 6076 (1995).

\bibitem{Sushkov2} O. P. Sushkov and V. N. Kotov, Phys. Rev. B {\bf 70}, 024503 (2004).

\bibitem{bazak} W. A. Atkinson, D. J. Bazak, and B. M. Andersen, Phys. Rev. Lett. {\bf 109}, 267004 (2012).

\bibitem{das14} T. Das, arXiv:1312.0544 (2014).

\bibitem{zhou2015} T. Zhou, Y. Gao, and J.-X. Zhu, Adv. Cond. Mat. Phys. {\bf 2015}, 652424 (2015).

\bibitem{Romer12} A. T. R\o mer, S. Graser, T. S. Nunner, P. J. Hirschfeld, and B. M. Andersen, Phys. Rev. B {\bf 86}, 054507 (2012).

\bibitem{EreminEPL02} I. Eremin, D. Manske, C. Joas, K.H. Bennemann Europhys. Lett. {\bf 58}, 871 (2002).
\bibitem{lee09} W.-C. Lee, S.-C. Zhang, and C. Wu, Phys. Rev. Lett. {\bf 102}, 217002 (2009).

\bibitem{platt12} C. Platt, R. Thomale, C. Honerkamp, S.-C. Zhang, and W. Hanke, Phys. Rev. B {\bf 85}, 180502(R) (2012).

\bibitem{maiti13} S. Maiti and A. V. Chubukov, Phys. Rev B {\bf 87}, 144511 (2013).

\bibitem{Hinojosa} A. Hinojosa, R. M. Fernandes, A. V. Chubukov, Phys. Rev. Lett. {\bf 113}, 167001 (2014).

\bibitem{Senegal} D. S\'en\'echal, P.-L. Lavertu, M.-A. Marois, and A.-M. S. Tremblay, Phys. Rev. Lett. {\bf 94}, 156404 (2005).


 \end{thebibliography}
\end{document}